\documentclass[twocolumn]{aastex631}
\usepackage{amsmath}
\usepackage{xcolor}
\usepackage{etoolbox}

\newcommand{\lisarates}{\texttt{lisarates}}

\newcommand{\msun}{\,M_\odot}

\newcommand{\Mc}{\mathcal{M}}
\newcommand{\Mtot}{M_{\rm tot}}

\newcommand{\missionYr}{4}

\newcommand{\snrThr}{8}
\newcommand{\snrThrHi}{15}
\newcommand{\ForecastHubble}{67.74}
\newcommand{\ForecastOmegaM}{0.3089}
\newcommand{\OfficialMergerThirty}{55^{+19}_{-9.8}\%}

\newcommand{\OfficialMergerTen}{39^{+22}_{-11}\%}
\newcommand{\NThreeBaseEight}{4.3^{+3.6}_{-2.0}}
\newcommand{\NThreeBaseFifteen}{0.63^{+0.54}_{-0.29}}
\newcommand{\NOfficialBaseEight}{6.6^{+5.5}_{-2.2}}
\newcommand{\NOfficialBaseFifteen}{0.91^{+0.64}_{-0.30}}
\newcommand{\NNtwoBaseEight}{5.7^{+4.9}_{-1.9}}
\newcommand{\NNthreeBaseEight}{5.9^{+6.6}_{-1.9}}
\newcommand{\NStwoBaseEight}{6.7^{+12}_{-2.6}}
\newcommand{\NSthreeBaseEight}{7.4^{+37}_{-3.2}}
\newcommand{\NNtwoExtEight}{5.9^{+9.5}_{-2.2}}
\newcommand{\NNthreeExtEight}{6.2^{+24}_{-2.4}}
\newcommand{\NStwoExtEight}{8^{+65}_{-3.9}}
\newcommand{\NSthreeExtEight}{14^{+3500}_{-9.4}}
\newcommand{\NTailWideEight}{6.9^{+43}_{-3}}
\newcommand{\NTailWideFifteen}{1.1^{+16}_{-0.49}}
\newcommand{\NNtwoAddedEight}{0.22^{+4.5}_{-0.21}}
\newcommand{\NNthreeAddedEight}{0.38^{+18}_{-0.38}}
\newcommand{\LocalRateNtwo}{19.3^{+5.2}_{-4.2}}
\newcommand{\MergerThirtyNtwo}{54^{+26}_{-11}\%}
\newcommand{\LocalRateNthree}{19.3^{+5.1}_{-4.1}}
\newcommand{\MergerThirtyNthree}{56^{+35}_{-13}\%}
\newcommand{\LocalRateStwo}{19.8^{+5.4}_{-4.2}}
\newcommand{\MergerThirtyStwo}{66^{+30}_{-22}\%}
\newcommand{\LocalRateSthree}{20.0^{+5.3}_{-4.3}}
\newcommand{\MergerThirtySthree}{80^{+20}_{-36}\%}
\newcommand{\MergerThirtyTailWide}{61^{+34}_{-16}\%}
\newcommand{\TailWideOutsideCoverage}{70^{+28}_{-44}\%}
\newcommand{\ThirdPeakNthree}{56^{+12}_{-9.1}}
\newcommand{\ThirdPeakSthree}{60^{+9.7}_{-11}}

\shorttitle{A quiet night in the millihertz sky?}
\shortauthors{Maga\~na Hernandez \& Palmese}

\graphicspath{{./}{figures/}}

\begin{document}

\title{A Quiet Night in the Millihertz Sky? LISA Forecasts from LIGO/Virgo/KAGRA Populations}

\correspondingauthor{Ignacio Maga\~na Hernandez} \email{imhernan@andrew.cmu.edu}

\author[0000-0003-2362-0459]{Ignacio Maga\~na Hernandez} \affiliation{McWilliams Center for Cosmology, Department of Physics, Carnegie Mellon University, Pittsburgh, PA 15213, USA}

\author[0000-0002-6011-0530]{Antonella Palmese} \affiliation{McWilliams Center for Cosmology, Department of Physics, Carnegie Mellon University, Pittsburgh, PA 15213, USA}

\begin{abstract}
The high-mass black hole population inferred from ground-based observations can substantially affect the number of binaries detectable by the Laser Interferometer Space Antenna (LISA). We use population constraints based on data from the latest LIGO/Virgo/KAGRA Gravitational-Wave Transient Catalog (GWTC-5.0), as well as constraints from the O3 search for intermediate-mass black hole binaries, to forecast the number and masses of binary black holes detectable by LISA. Using the released GWTC-5.0 population posterior, we obtain $\NOfficialBaseEight$ expected detections with signal-to-noise ratio $>8$ in \missionYr{} years (median and $90\%$ credible interval). The alternative population models we consider use a broken power law with two or three Gaussian peaks for either the primary-mass distribution or a common mass distribution for both black holes. For the three-peak model with the common mass distribution, extrapolating from $300$ to $1000\msun$ roughly doubles the median expected number of LISA detections to $\sim14$. The upper end of its $90\%$ credible interval rises from tens to thousands of expected detections, although approximations may limit the reliability of the most extreme counts. For the reference GWTC-5.0 population, $\OfficialMergerTen$ of the binaries detectable by LISA merge during the mission or within $10$ years after its end and are potential multiband sources. Essentially all of these mergers would be detectable by a reference Cosmic Explorer configuration if it is operating at merger. Ground-based measurements of the high-mass population can therefore inform both LISA forecasts and the prospects for observing the same binaries through inspiral and merger.
\end{abstract}


\section{Introduction}
\label{sec:intro}

The Laser Interferometer Space Antenna \citep[LISA;][]{Colpi:2024LISARedBook} will observe the millihertz inspiral of binary black holes (BBHs). Some of these systems will merge in the frequency band of the ground-based LIGO--Virgo--KAGRA (LVK) detectors \citep{Aasi:2015AdvancedLIGO,Acernese:2015AdvancedVirgo,Akutsu:2021KAGRA} within years to decades of their LISA observation, allowing the same binary to be observed by LISA during its inspiral and by ground-based detectors near merger \citep{Sesana:2016,Gerosa:2019Multiband}.

At millihertz frequencies, measurements of orbital eccentricity can help distinguish how black hole binaries formed, before gravitational radiation further circularizes their orbits \citep{Breivik:2016FormationChannels,Nishizawa:2017FormationEccentricity,Cutler:2019MultibandScience}. For systems observed in both bands, combining LISA's early inspiral with ground-based observations near merger can improve measurements of binary parameters \citep{Vitale:2016Multiband,Toubiana:2022GWTC3LISA} and strengthen tests of general relativity \citep{Toubiana:2020LISATestsGR,Datta:2021MultibandGR}. These possibilities motivate forecasts of both the number and masses of binary black holes accessible to LISA.

Forecasts based on the third Gravitational-Wave Transient Catalog (GWTC-3; \citealt{Abbott:2023GWTC3cat,Abbott:2021GWTC3pop}) predict only a few individually resolved BBHs in a four-year mission at signal-to-noise ratio (SNR) $\rho>8$ \citep{Buscicchio:2025SOBBH}. Searches without prior source information may require higher SNR thresholds; \citet{Moore:2019Quiet} forecast fewer than one multiband early warning within a ten-year merger window for most of their modeled populations. A small catalog limits population measurements, but massive BBHs whose inspirals lie in LISA's sensitive band can be detected over a larger volume than lighter systems \citep{Gerosa:2019Multiband,Wagg:2022LEGWORK}. They can therefore make up a larger fraction of LISA detections than of the intrinsic population.

The populations inferred from GWTC-3 \citep{Abbott:2021GWTC3pop}, GWTC-4.0 \citep{Abbott:2025GWTC4pop}, and GWTC-5.0 \citep{Abbott:2025GWTC5pop} contain several features in the source-frame primary mass distribution, including a peak near $10\msun$ and a change in slope around $35\msun$ \citep{Abbott:2025GWTC5pop}. Other analyses find model-dependent structure at higher masses \citep{TiwariFairhurst:2021Structure,Pierra:2026HeavyBH}, including tentative evidence for a subdominant component near $60-70\msun$ \citep{Abbott:2021GWTC2pop,MaganaHernandez:2024Bump,MaganaHernandez:2025GP}. Beyond features in the primary mass distribution, detectability will also depend on the secondary mass distribution and on how the black holes are paired \citep{FishbachHolz:2020Pairing,Farah:2024TwoOfAKind}.

GW190521 and GW231123 probe the high-mass black hole population in and around the range where pair instability can suppress black-hole formation through stellar collapse \citep{Farmer:2019MindTheGap,Abbott:2020GW190521,Abbott:2020GW190521Implications,Abac:2025GW231123}. Dedicated searches for intermediate-mass black hole (IMBH) binaries have also constrained the rates of these rare systems \citep{Abbott:2019IMBHO1O2,Abbott:2022IMBHO3}. At intermediate masses, forecasts identify binaries that could be observed by both LISA and next-generation ground-based detectors \citep{Mangiagli:2019AboveGap,Jani:2019IMBHMultiband,FragioneLoeb:2023IMBHHistory}.

We first use the released GWTC-3 and GWTC-5.0 population posteriors to update the expected LISA population and compare with earlier forecasts. We then consider alternative fits to GWTC-5.0, with two or three mass peaks, using either a primary-mass distribution with a conditional mass-ratio distribution or a common mass distribution for both black hole masses. A common underlying mass distribution is physically motivated when both black holes are drawn from the same population, as can occur in dynamically assembled binaries \citep{Rodriguez:2016ClusterBBH,Antonini:2023ClusterMasses}. Analyses of GWTC-3 find that shared mass features in both components are consistent with the data \citep{Farah:2024TwoOfAKind}. The pairing weight allows correlations between the component masses \citep{FishbachHolz:2020Pairing}, as expected when dynamical interactions preferentially pair massive black holes \citep{Torniamenti:2024Pairing}. Comparing these models tests how the LISA forecasts depend on the assumed mass distributions and pairing prescriptions. To assess the contribution of binaries beyond the fitted mass range, we extrapolate each model from its upper primary-mass limit of $300\msun$ to $1000\msun$, since LISA sensitivity is greater in this range than at lower masses.

For all the LISA forecasts in this work, we adopt a fiducial, flat $\Lambda$CDM cosmology with $H_0=\ForecastHubble\,\mathrm{km\,s^{-1}\,Mpc^{-1}}$ and $\Omega_{\rm m}=\ForecastOmegaM$, following \citet{Planck:2015Cosmo}. Section~\ref{sec:models} describes the population models and LISA selection. Section~\ref{sec:results} presents the catalog baselines. Sections~\ref{sec:results:models} and \ref{sec:results:highmass} examine the alternative mass models and the less restrictive high-mass extrapolation. Section~\ref{sec:multiband} considers multiband sources, and Section~\ref{sec:discussion} discusses what the forecasts imply for the high-mass population and how they depend on the assumed mass distribution and merger-rate history.

\section{Population Models and LISA}
\label{sec:models}

\subsection{Baseline Populations}
\label{sec:popmodels}

We use the released GWTC-3 and GWTC-5.0 population posteriors \citep{LVK:2024GWTC3PopulationData,LVK:2026GWTC5PopulationData} for our reference forecasts. The GWTC-5.0 \textsc{Brokenpowerlaw+2Peaks} model describes the primary mass with a broken power law and two Gaussian peaks, a smooth lower cutoff, and a fixed upper limit of $300\msun$ \citep{Abbott:2025GWTC5pop}. Its inferred parameters include the power-law slopes and break location, the peak locations, widths and weights, and the low-mass cutoff and smoothing scale. The GWTC-3 \textsc{Powerlaw+Peak} model instead has one power-law slope and one Gaussian peak, and infers the upper mass cutoff \citep{TalbotThrane:2018Mass,Abbott:2021GWTC3pop}. In both models, the companion distribution at fixed primary mass follows a power law in $q=m_2/m_1$, with a low-mass taper and normalization over the allowed secondary masses.

\begin{figure*}[t]
\centering
\includegraphics[width=\textwidth]{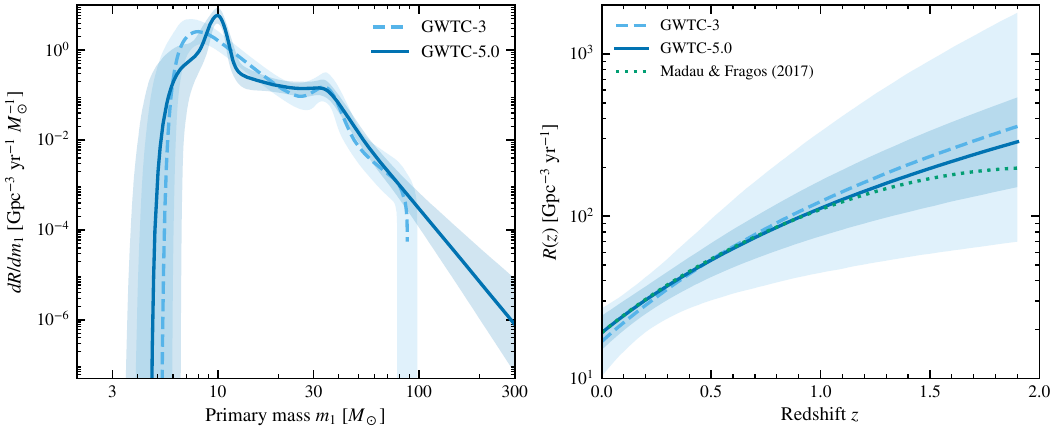}
\caption{
The released catalog populations used for the baseline forecasts. \emph{Left:} local merger rate per unit source-frame primary mass for the GWTC-3 \textsc{Powerlaw+Peak} and GWTC-5.0 \textsc{Brokenpowerlaw+2Peaks} posteriors \citep{Abbott:2021GWTC3pop,Abbott:2025GWTC5pop}. The GWTC-3 upper mass cutoff is inferred and differs between posterior samples, while the GWTC-5.0 upper cutoff is fixed at $300\msun$. \emph{Right:} their inferred merger-rate evolution. The dotted curve shows the fixed star-formation-history shape of \citet{Madau:2017SFH}, normalized to the median local GWTC-5.0 rate. Curves and shaded bands show the median and central $90\%$ credible interval at each mass or redshift.}
\label{fig:popinputs}
\end{figure*}

Figure~\ref{fig:popinputs} shows the inferred mass distributions and merger-rate evolution for both GWTC-3 and GWTC-5.0. We assume that the mass distribution does not evolve with redshift \citep{Abbott:2025GWTC5pop}. For each released catalog population, we use its inferred rate evolution, $\mathcal R(z)=\mathcal R_0(1+z)^\kappa$ \citep{Fishbach:2018Redshift,Abbott:2021GWTC3pop,Abbott:2025GWTC5pop}. Each posterior sample contains one complete set of population parameters, including the mass and spin distributions, local merger rate, and redshift exponent. We use these parameters together when calculating a forecast. Appendix~\ref{app:popmodels} defines the population densities.

\subsection{Alternative Mass Models}
\label{sec:fitmodels}

We fit four mass models to a common set of 259 GWTC-5.0 binary black hole events, including GW190521 and GW231123 \citep{LVK:2026GWTC5Intro,LVK:2026GWTC5Results,Abbott:2025GWTC5pop}. The primary-mass models combine a broken power law (BPL) with two or three Gaussian peaks (2P or 3P), and are denoted \texttt{BPL2P} and \texttt{BPL3P}. The companion is described by a conditional mass-ratio distribution. The symmetric models, \texttt{symBPL2P} and \texttt{symBPL3P}, use the same underlying distribution for both component masses before pairing, with pairs weighted by $q^\beta$ \citep{FishbachHolz:2020Pairing,Farah:2024TwoOfAKind}. Writing the underlying mass distribution as $B_K$, the primary-mass and symmetric models give $p(m_1,q)=B_K(m_1)p(q|m_1)$ and $p(m_1,q)\propto m_1B_K(m_1)B_K(qm_1)q^\beta$, respectively. We express both models in $(m_1,q)$. For the symmetric models, changing variables from $(m_1,m_2)$ introduces the factor $m_1$. Here $K=2$ or $3$, $q=m_2/m_1$, and $m_1\geq m_2$; Appendix~\ref{app:fitinference} defines the normalizations and low-mass tapers.

The two-peak fits describe the structure near $10$ and $30\msun$. For the three-peak fits, we allow the additional peak's mean to lie between $45$ and $90\msun$ and fit its width and abundance along with the rest of the population.  All four fits restrict component masses to $3$--$300\msun$, allow the location and width of the smooth low-mass cutoff to vary, and share a Gaussian effective-spin distribution and power-law merger-rate evolution (Appendix~\ref{app:rateinference}).

\texttt{BPL2P} uses the same mass-model family as the released LVK \textsc{Brokenpowerlaw+2Peaks} posterior, but we infer its parameters independently. We use the released posterior for the catalog baseline forecasts. The additional peak is motivated by the reported feature near $70\msun$ \citep{MaganaHernandez:2024Bump,MaganaHernandez:2025GP}, and we are specifically interested in its addition to explore whether it could significantly contribute to the LISA detectable population. We note that an additional high mass peak has also been used for LVK cosmological analyses \citep{Abac:2026GWTC5Cosmo}. These fits compare mass models with two or three peaks and different prescriptions for pairing the component masses.

The observed number of mergers and the LVK detection efficiency determine the local merger rate for each fitted population. Populations that are harder to detect require a higher rate to explain the same catalog. We propagate both the uncertainty in this rate and its dependence on the mass, spin, and redshift parameters into the LISA forecasts \citep{Mandel:2019Selection,ThraneTalbot:2019Inference}.

We fit both the primary-mass and symmetric models to GWTC-5.0 using hierarchical population inference, jointly inferring the mass, spin, and redshift parameters \citep{Mandel:2019Selection,ThraneTalbot:2019Inference}. We evaluate the population likelihood using samples from the event posteriors and account for selection effects using simulated signals recovered by the ground-based searches \citep{Farr:2019Selection,LVK:2026GWTC5Sensitivity,Essick:2025O4Injections}. To limit numerical error from these finite samples, we require the estimated variance of the log likelihood to satisfy $\widehat\sigma^2_{\ln\mathcal L}<1$ \citep{TalbotGolomb:2023Likelihood,Abbott:2025GWTC4pop}. Parameter combinations that fail this requirement are excluded from the fit. This also restricts which populations can be explored. Appendix~\ref{app:rateinference} discusses this limitation.

\subsection{Extrapolation to Higher Masses}
\label{sec:continuations}

For each fitted model, we extend the primary-mass range from $300$ to $1000\msun$ using the same mass dependence and pairing prescription. The merger rate assigned to binaries with $m_1\leq300\msun$ is unchanged. Binaries with larger primary masses (while their companions may lie below or above $300\msun$) add to that rate. Appendix~\ref{appendix:extend} gives the corresponding rate expressions.

Forecasts using the fitted power-law merger-rate evolution are restricted to $0\leq z\leq1.9$. The added massive binaries can be detectable at higher redshifts, so for the extrapolations we use a merger-rate history proportional to the star-formation history of \citet{Madau:2017SFH}, scaled to match the inferred local merger rate below $300\msun$ at $z=0$ (Appendix~\ref{appendix:extend}). We neglect delays between binary formation and merger, which can change the merger-rate evolution with redshift \citep{FishbachKalogera:2021TimeDelays}.

We also calculate forecasts restricted to $m_1\leq300\msun$ with this same redshift evolution. Comparing these with the extrapolations to $m_1\leq1000\msun$ isolates the contribution from binaries with $m_1>300\msun$. We repeat these calculations across the inferred population parameters and local rates, so that the spread in their predictions gives the quoted credible intervals.

To allow a different abundance of binaries above $300\msun$, we also consider a free-slope extrapolation of \texttt{BPL2P}. Its mass-distribution slope above $300\msun$ varies independently of the fitted slope. We set $d\mathcal R/dm_1\propto m_1^{-\alpha}$ up to $1000\msun$, with $\alpha\sim U[-2,8]$, and match the rate density at $300\msun$ to the fitted model. This range allows both rising and falling high-mass distributions. We keep the same conditional mass-ratio prescription and star-formation-history evolution. We use the O3 IMBH search sensitivity \citep{Abbott:2022IMBHO3} to weight the allowed slopes by the probability of no detections with primary mass above $300\msun$ within its sensitivity coverage. This update changes the slope distribution conditional on each fitted population and rate, but it does not refit GWTC-5.0 (see Appendix~\ref{app:o3}).

\subsection{LISA Detections}
\label{sec:forecast}

We adopt a \missionYr-yr observing interval and classify binaries with mission-integrated signal-to-noise ratio $\rho>\snrThr$ as resolved. We also report forecasts for $\rho>\snrThrHi$, motivated by \citet{Moore:2019Quiet}, who estimated thresholds for searches without prior source information for binaries with component masses of $5$--$50\msun$ and merger times up to $100$ years after mission start. We use the higher threshold as a comparison for all our populations, including the higher-mass extrapolations. We do not calculate a search threshold specific to each mass or merger time.

We integrate the signal over the frequencies traversed during the mission, or until merger if it occurs first. We calculate the SNR using two effective channels. The noise model includes the instrumental contribution of \citet{Babak:2021SNR} and the foreground from unresolved Galactic double white dwarfs of \citet{Robson:2019LISASens}. We average the response over sky position and polarization and integrate over isotropic inclinations. We assume quasi-circular inspirals and use the same noise spectrum for every population. Appendix~\ref{app:noise} specifies the noise spectrum and the conversion from binary strain to the two-channel SNR. For \texttt{symBPL3P}, which permits the largest numbers of massive binaries, we also compare the total binary black hole background predicted by this model with the assumed noise (Appendix~\ref{app:foreground}). This tests whether neglecting that background could affect our forecasts.

The forecast includes binaries observable during the mission even if they merge long after it ends. For a specified population, we integrate over masses, redshifts, and merger times, weighting each contribution by the fraction of inclinations that exceed the SNR threshold. The mission duration sets the observed frequency interval \citep{Babak:2023SGWB,Buscicchio:2025SOBBH}. Repeating the integral over the inferred population parameters and rates gives the distribution of expected counts. Appendix~\ref{app:sampling} shows the integral and its limits.

The SNR estimation involves comparing the signal with the detector noise at each frequency. At leading order, the inspiral strain amplitude in the frequency domain scales as $|\widetilde h(f)|\propto \Mc_{\rm det}^{5/6}f^{-7/6}/d_L$, where $\Mc_{\rm det}=(1+z)\Mc$ is the detector-frame chirp mass and $\Mc=(m_1m_2)^{3/5}/(m_1+m_2)^{1/5}$ is the source-frame chirp mass \citep{CutlerFlanagan:1994Parameters}. Here $d_L$ is the luminosity distance and $f$ is the observed frequency. The accumulated SNR also depends on the portion of the inspiral observed and the detector noise over that interval. For inspirals within LISA's sensitive band, the larger amplitude of massive binaries can increase their detectable volume. They can therefore make up a larger fraction of LISA detections than they are in the astrophysical merger population \citep{Gerosa:2019Multiband,Wagg:2022LEGWORK}.

\begin{figure*}[t]
\centering
\includegraphics[width=\textwidth]{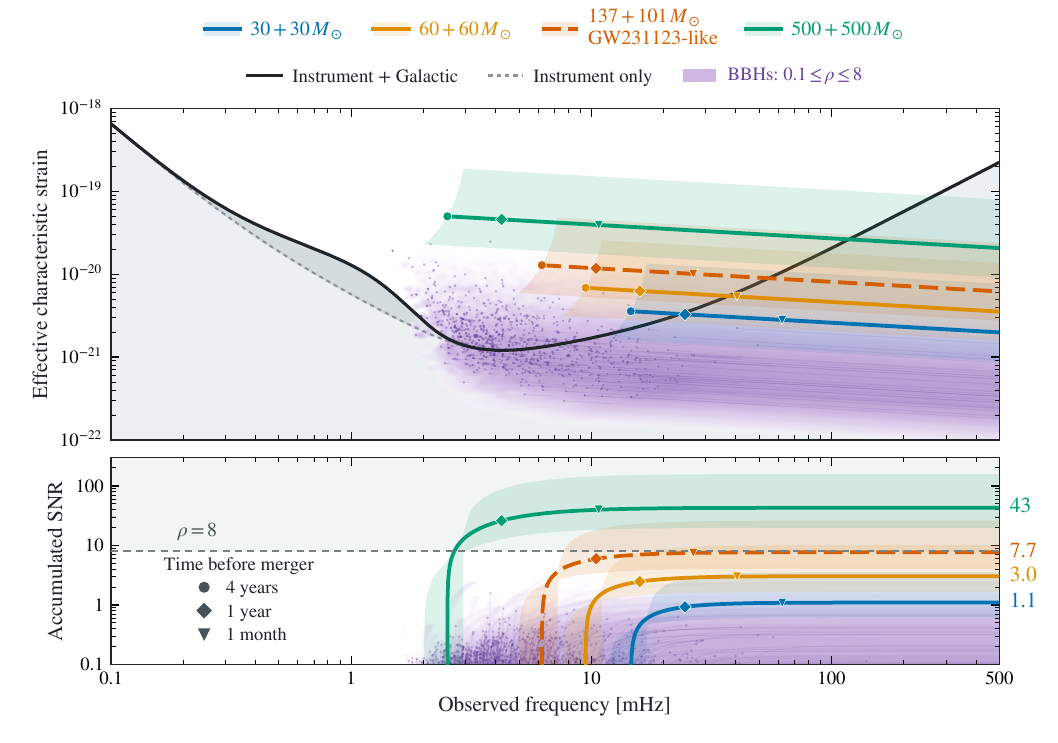}
\caption{LISA signals from four binaries with the source-frame component masses shown in the legend, observed during their final four years before merger. The GW231123-like binary has masses $137+101\msun$ \citep{Abac:2025GW231123}. The upper panel shows effective characteristic strain and LISA sensitivity with and without the Galactic double white dwarf foreground. The lower panel shows accumulated two-channel SNR, including this foreground. Bold curves correspond to $z=0.40$, and colored bands span $0.1\leq z\leq1.0$ at fixed source-frame masses. These signals are averaged over sky position and polarization, with the root-mean-square response over isotropic inclinations. Symbols mark the times before merger listed in the legend; numbers on the right give the final SNRs at $z=0.40$. The horizontal dashed line marks the detection threshold, $\rho=8$. Purple tracks show individual BBHs with four-year SNRs $0.1\leq\rho\leq8$ from one \texttt{BPL2P} population extrapolated to $1000\msun$. Purple points mark observation starts in the upper panel and endpoints in the lower panel. Purple shading indicates the relative density of individual tracks, not a summed BBH background. See Appendix~\ref{app:snr} for the strain normalization and population-sample details.}
\label{fig:lisasensitivity}
\end{figure*}

Figure~\ref{fig:lisasensitivity} compares four reference mass pairs over $0.1\leq z\leq1.0$, observed during their final four years before merger. In these examples, larger masses produce stronger signals and shift the observed inspiral toward LISA's most sensitive frequencies. The purple tracks show the strains and accumulated SNRs of BBHs that remain below our detection threshold over the mission.

For a GW231123-like binary with fixed source-frame masses of $137+101\msun$, the final four years before merger give $\rho\simeq7.7$ at $z=0.40$, using the angular averaging in Figure~\ref{fig:lisasensitivity}. At the lower and upper limits of the reported $90\%$ redshift interval, $z=0.15$ and $0.67$ \citep{Abac:2025GW231123}, we obtain $\rho\simeq18$ and $5.2$, respectively. With these assumptions, $\rho=8$ at $z\simeq0.38$, so that depending on the exact redshift a merger of this kind could be detectable by LISA. This comparison varies redshift alone and does not propagate the event's correlated uncertainties in masses and orientation.

\section{Results}
\label{sec:results}

In what follows, we report our expectations for LISA detections of binary black holes. We quote posterior medians with lower and upper errors spanning the central $90\%$ credible interval (CI). For forecast counts, these intervals describe uncertainty in the expected number of detections from the population parameters and merger rate. An actual catalog has an additional Poisson fluctuation about that expectation, which is not included in the quoted intervals. Numerical integration error is assessed separately in Appendix~\ref{app:sampling}.

\subsection{Baseline Population Forecasts}
\label{sec:results:floor}

For the released GWTC-5.0 population, the median expected number of resolved binaries is $\NOfficialBaseEight$ at $\rho>\snrThr$ in a \missionYr-yr mission. Using the GWTC-3 population posterior, we obtain $\NThreeBaseEight$. In the GWTC-5.0 forecast, binaries with $100\msun<m_1\leq300\msun$ contribute $1.4^{+5.0}_{-1.1}$ expected detections, or $21^{+33}_{-15}\%$ of the total. Binaries with $m_1\leq100\msun$ contribute $5.0^{+1.8}_{-1.4}$, compared with $\NThreeBaseEight$ for GWTC-3. The higher GWTC-5.0 forecast is therefore driven mainly by binaries with primary masses between $100$ and $300\msun$, while the forecasts below $100\msun$ remain similar. \citet{Buscicchio:2025SOBBH} report a mean of about five detections for a four-year mission at $\rho>8$. Appendix~\ref{app:snr} specifies the SNR convention used here and its relation to that work.

After averaging over population uncertainty and Poisson count fluctuations, the released GWTC-5.0 population gives a probability of about $0.3\%$ that LISA detects no binaries with $\rho>\snrThr$ in four years. At the higher threshold $\rho>\snrThrHi$, the expected counts fall to $\NOfficialBaseFifteen$ for GWTC-5.0 and $\NThreeBaseFifteen$ for GWTC-3.

\subsection{Alternative Mass Model Forecasts}
\label{sec:results:models}

\subsubsection{Forecasts to 300 $M_\odot$}

Within the fitted mass range and using the inferred power-law merger-rate evolution, \texttt{BPL2P} and \texttt{BPL3P} predict $\NNtwoBaseEight$ and $\NNthreeBaseEight$ detections at $\rho>\snrThr$ in four years, respectively. \texttt{symBPL2P} and \texttt{symBPL3P} predict $\NStwoBaseEight$ and $\NSthreeBaseEight$, with broader credible intervals (See Table~\ref{tab:summary}).

The inferred local merger rates on the fitted mass range are $\LocalRateNtwo$, $\LocalRateNthree$, $\LocalRateStwo$, and $\LocalRateSthree\,\mathrm{Gpc}^{-3}\mathrm{yr}^{-1}$ for \texttt{BPL2P}, \texttt{BPL3P}, \texttt{symBPL2P}, and \texttt{symBPL3P}, respectively. The local rates are similar across the four models. Their LISA forecasts differ because the mass distributions allow different numbers of binaries with two massive components. In the symmetric mass models, the companion distribution depends on the same high-mass structure as the primary, as well as on the pairing weight, and it therefore may produce more LISA detections than the other mass models, which instead describe the companion through a conditional power law in mass ratio.

In \texttt{BPL3P} and \texttt{symBPL3P}, the fitted additional peak has mean $\ThirdPeakNthree\msun$ and $\ThirdPeakSthree\msun$, respectively. Within the fitted mass range, the third Gaussian accounts for $0.45^{+0.81}_{-0.39}\%$ and $0.52^{+0.65}_{-0.42}\%$ of primary black holes in the intrinsic population, respectively. Its location, width, and abundance are inferred together with the rest of the population. The count intervals overlap within the fitted mass range. The fits also change the continuum, pairing, and low-mass cutoffs, so the forecast differences cannot be attributed to the additional peak alone. The symmetric three-peak fit permits larger expected counts, with the upper end of its $90\%$ credible interval exceeding $\sim40$ detections.

\begin{figure*}[t]
\centering
\includegraphics[width=\textwidth]{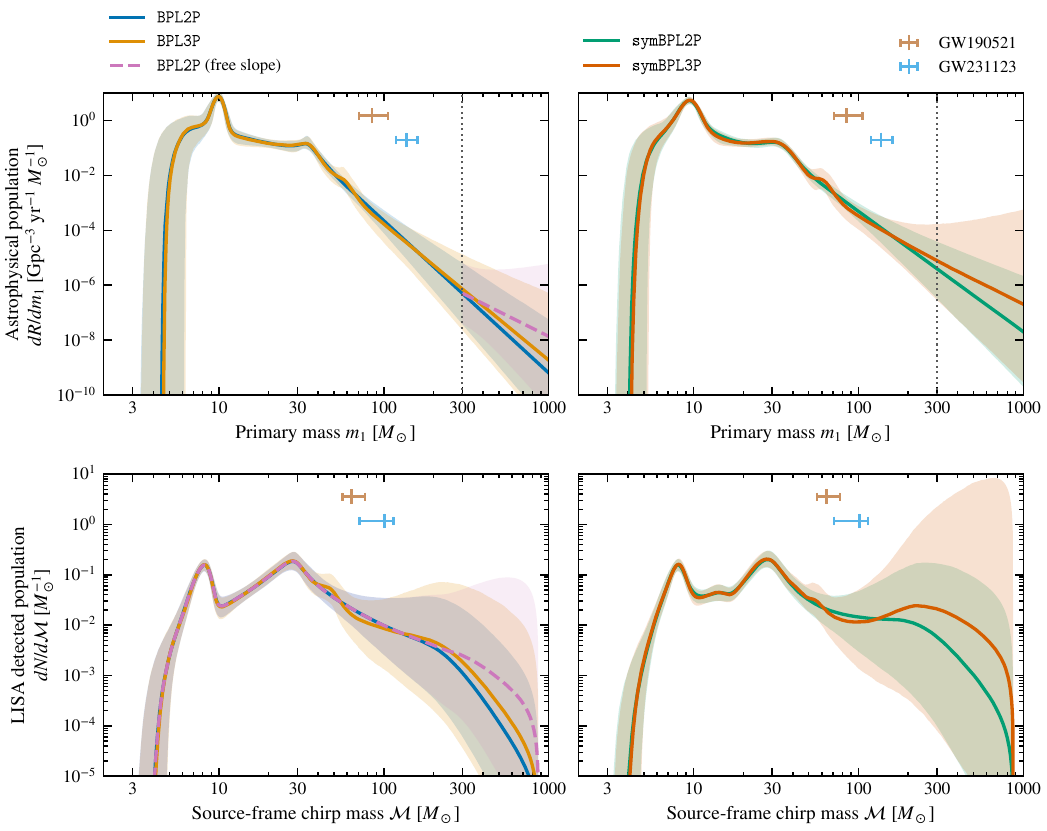}
\caption{
Mass distributions extrapolated to $1000\msun$. The primary-mass models (left), \texttt{BPL2P} and \texttt{BPL3P}, combine a broken power law with two or three Gaussian peaks. Their symmetric counterparts (right), \texttt{symBPL2P} and \texttt{symBPL3P}, use a common mass distribution for both components before pairing. \emph{Upper panels:} the astrophysical local merger rate per unit primary mass. The dashed magenta curve shows the free-slope extrapolation of \texttt{BPL2P}. The dotted vertical line marks the fitted upper mass limit. \emph{Lower panels:} the expected LISA detections per unit source-frame chirp mass for the four fitted-slope extrapolations and the free-slope \texttt{BPL2P} extrapolation, with $\rho>\snrThr$ in \missionYr{} years and the Madau--Fragos merger-rate history. Curves and shading show medians and $90\%$ CI at each mass. The distributions are not rescaled to unit area. Horizontal bars show the central $90\%$ source-frame $m_1$ (upper) and $\Mc$ (lower) intervals for GW190521 and GW231123 \citep{Abbott:2020GW190521,Abac:2025GW231123}, with vertical ticks at the medians; bar heights are arbitrary.}
\label{fig:massmodels}
\end{figure*}

\subsubsection{Extrapolation to 1000 $M_\odot$}

The GWTC-5.0 population extends to primary masses above $100\msun$, with GW231123 providing a key observation at its high-mass end \citep{Abbott:2025GWTC5pop,Abac:2025GW231123}. Beyond component masses of approximately $200\msun$, the predicted abundance increasingly depends on the assumed high-mass shape of the population model \citep{Abbott:2025GWTC5pop}. For equal-mass binaries and an inclination-averaged response, the four-year $\rho=\snrThr$ horizon rises from approximately $z=0.1$ at source-frame total mass $\Mtot=100\msun$ to $z=4$ at $\Mtot=1000\msun$ (Figure~\ref{fig:noisehorizon}). For these horizon curves, we choose the gravitational-wave frequency at the start of the mission that gives the largest four-year SNR. The dependence of the SNR on time to merger is described by \citet{Gerosa:2019Multiband}.

The extrapolated \texttt{BPL2P} and \texttt{BPL3P} models give $\NNtwoExtEight$ and $\NNthreeExtEight$ expected detections, respectively. Their median counts remain close to the forecasts restricted to $300\msun$ with the same rate evolution, although the uncertainty increases. Binaries with $m_1>300\msun$ contribute $\NNtwoAddedEight$ and $\NNthreeAddedEight$ expected detections, respectively.

\begin{figure*}[t]
\centering
\includegraphics[width=\textwidth]{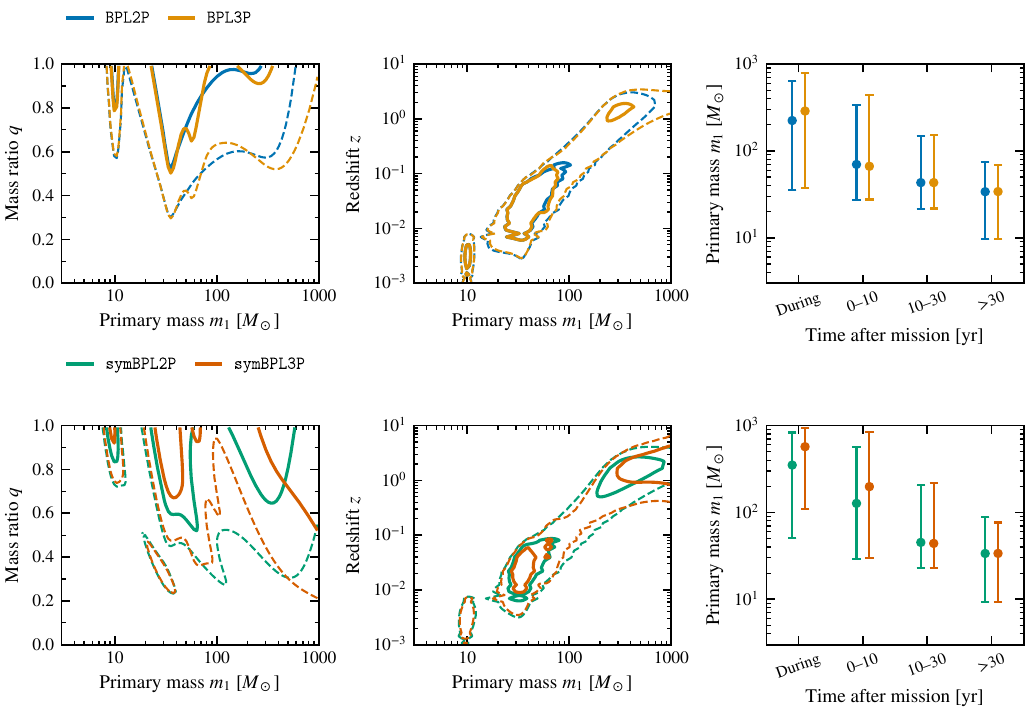}
\caption{Resolved LISA binary black hole population for the four mass models extrapolated to $1000\msun$, with $\rho>\snrThr$ in \missionYr{} years and the Madau--Fragos merger-rate history \citep{Madau:2017SFH}. The upper row compares the primary-mass models \texttt{BPL2P} and \texttt{BPL3P}; the lower row compares the symmetric models \texttt{symBPL2P} and \texttt{symBPL3P}. All masses are in the source frame. \emph{Left and middle:} joint distributions of primary mass with mass ratio and redshift. Solid and dashed contours enclose $50\%$ and $90\%$ of the detected-source probability, with density defined in the displayed coordinates: logarithmic in primary mass and redshift, and linear in mass ratio. \emph{Right:} median primary mass and central $90\%$ range of detected binaries in each merger-time window. ``During'' means during the LISA mission; the other labels give time intervals measured from mission end. We average the normalized detected distributions over the population posterior. The contours and mass ranges describe the spread of source properties among detectable binaries, not uncertainties in an individual binary's measured parameters.}
\label{fig:correlations}
\end{figure*}

The \texttt{symBPL2P} and \texttt{symBPL3P} extrapolations give a larger change, with $\NStwoExtEight$ and $\NSthreeExtEight$ expected detections, respectively. The high-mass slopes of the broken power law in the primary-mass distribution are $\alpha_2=5.5^{+1.5}_{-1.6}$ and $5.0^{+2.1}_{-2.3}$ for \texttt{BPL2P} and \texttt{BPL3P}. For the two- and three-peak symmetric models, the slopes of the common component-mass distribution before pairing are $2.7^{+1.2}_{-1.1}$ and $1.9^{+1.7}_{-2.1}$, respectively. These models favor shallower high-mass distributions; the three-peak model also allows a rising distribution within its $90\%$ credible interval. Extrapolating those shapes can add many binaries with two massive components, to which LISA is particularly sensitive. This is especially important at the upper end of the \texttt{symBPL3P} forecast interval (Figure~\ref{fig:massmodels}). Figure~\ref{fig:correlations} shows how primary mass is related to mass ratio, redshift, and merger time among the detectable binaries. At high primary mass, the $90\%$ contours extend to lower mass ratios for the symmetric models, particularly \texttt{symBPL3P}. Higher-mass detections extend to larger redshifts in all four models, consistent with the mass dependence of LISA's reach (Figure~\ref{fig:noisehorizon}). We discuss the merger-time dependence in Section~\ref{sec:multiband}.

\begin{figure}[htbp]
\centering
\includegraphics[width=\columnwidth]{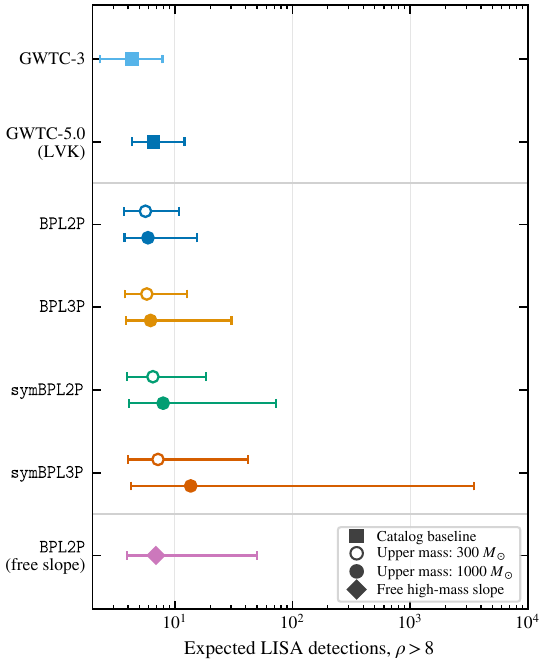}
\caption{Expected LISA detections with $\rho>\snrThr$ in \missionYr{} years. Squares denote forecasts from the released GWTC-3 and GWTC-5.0 posteriors. For each of our four GWTC-5.0 fits, open circles give the $300\msun$ calculation and filled circles its extrapolation to $1000\msun$, both with the same Madau--Fragos merger-rate history. The diamond denotes the free-slope extrapolation of \texttt{BPL2P}. Symbols and bars show medians and $90\%$ intervals from population and merger-rate uncertainty. Table~\ref{tab:summary} gives the numerical values and the $\rho>\snrThrHi$ comparison.}
\label{fig:countforest}
\end{figure}

The reported forecasts use the same instrumental noise and unresolved Galactic double-white-dwarf foreground. Adding the total binary black hole background predicted by \texttt{symBPL3P} to the noise reduces the expected counts for populations near the upper end of its forecast interval (Appendix~\ref{app:foreground}).

\subsection{Free High-Mass Slope}
\label{sec:results:highmass}

The free-slope calculation allows the abundance of binaries above $300\msun$ to differ from an extrapolation of the fitted power law. The two distributions meet at the same rate per unit primary mass at $300\msun$.

For \texttt{BPL2P}, the free-slope extrapolation gives $\NTailWideEight$ expected detections at $\rho>\snrThr$, compared with $\NNtwoExtEight$ when the fitted slope is retained. The median changes little, while the upper end of the $90\%$ credible interval increases. At $\rho>\snrThrHi$, the free-slope forecast falls to $\NTailWideFifteen$. The upper-left panel of Figure~\ref{fig:massmodels} compares the fitted-slope and free-slope mass distributions.

Before applying the O3 IMBH constraint, the added population predicts $(5.9^{+72}_{-5.5})\times10^{-5}$ O3 detections within the mass and mass-ratio region covered by the nonspinning injections used here. Weighting the slopes by the probability of this non-detection (see Appendix~\ref{app:o3}) therefore leaves the slope distribution and LISA forecast essentially unchanged. The count interval reflects the range of high-mass slopes we allow, together with uncertainty in the fitted population and merger rate.

The O3 sensitivity study includes binaries with total source-frame masses up to $800\msun$, but its sampled configurations have primary masses no larger than $400\msun$ \citep{Abbott:2022IMBHO3}. Here we use its nonspinning injections as a reference calculation. For the free-slope model, $\TailWideOutsideCoverage$ of the added LISA detections with $m_1>300\msun$ and $\rho>\snrThr$ lie outside the mass and mass-ratio region covered by these injections. This fraction accounts for the predicted primary masses and mass ratios, and it does not include spin-dependent sensitivity.

Figure~\ref{fig:countforest} summarizes the expected counts for the released GWTC-3 and GWTC-5.0 populations, the four alternative mass models, and the free-slope extrapolation. For each fitted mass model, it compares the extrapolations with the corresponding $300\msun$ calculations using the same merger-rate evolution with redshift.

\subsection{Multiband Implications}
\label{sec:multiband}

The time to coalescence sets when a LISA binary could be observed near merger by ground-based detectors. We count binaries with $t_{\rm c,det}\leq T_{\rm obs}+\Delta t$, where $t_{\rm c,det}$ is the observer-frame time to merger from mission start, $T_{\rm obs}=\missionYr$ yr is the mission duration, and $\Delta t$ is the  time after it ends. We report mergers during the mission and within $10$ or $30$ years after its end. We first report merger-time fractions, then assess ground-based detectability with a reference Cosmic Explorer (CE) configuration \citep{Reitze:2019CE,Evans:2021CEHorizon}. Within the $30$-year window, the fraction of LISA detections that merge is $\OfficialMergerThirty$ when assuming the baseline GWTC-5.0 population.

\begin{figure*}[t]
\centering
\includegraphics[width=\textwidth]{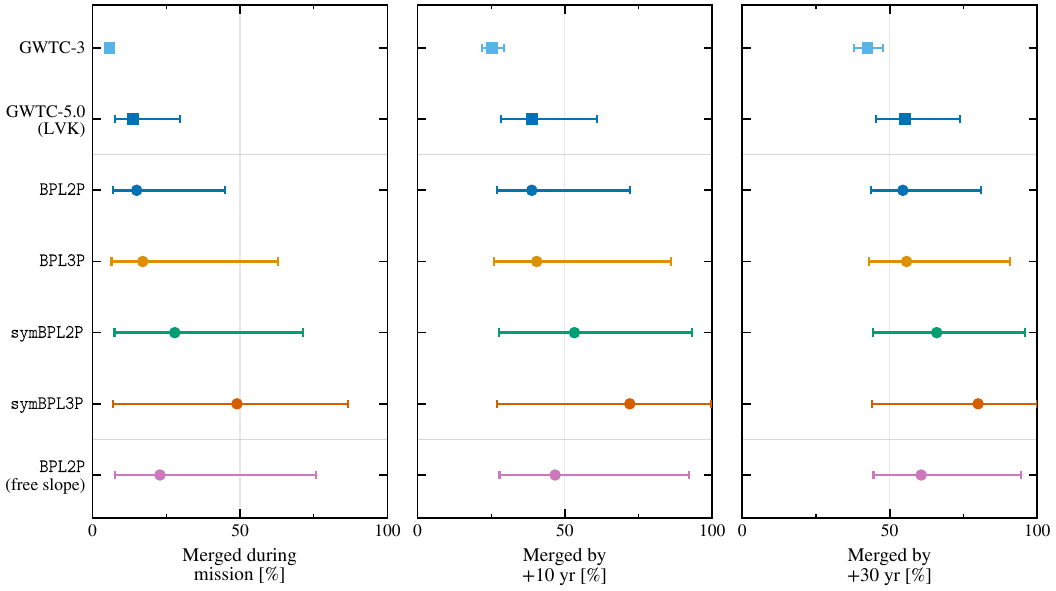}
\caption{Fractions of the $\rho>\snrThr$ LISA population that merge during the four-year mission or within 10 or 30 years after its end, from left to right. Each window includes mergers during the mission. Squares denote the released GWTC-3 and GWTC-5.0 posteriors; circles denote our four GWTC-5.0 fits extrapolated to $1000\msun$ with the Madau--Fragos merger-rate history. The free-slope row uses \texttt{BPL2P} with an independent slope above $300\msun$. Symbols and bars show medians and central $90\%$ credible intervals from population uncertainty. No ground-based detection requirement is imposed. Table~\ref{tab:multiband} gives the corresponding numerical results.}
\label{fig:mergertimes}
\end{figure*}

For extrapolations to $1000\msun$, the fractions merging during the mission or within $30$ years after its end are $\MergerThirtyNtwo$ for \texttt{BPL2P} and $\MergerThirtyNthree$ for \texttt{BPL3P}. They increase to $\MergerThirtyStwo$ and $\MergerThirtySthree$ for \texttt{symBPL2P} and \texttt{symBPL3P}, respectively. Allowing a free high-mass slope in \texttt{BPL2P} gives a fraction of $\MergerThirtyTailWide$. Figure~\ref{fig:mergertimes} compares the merger fractions during the mission and through 10 and 30 years after its end. The broad intervals, especially for the extrapolated models, reflect the uncertain contribution of massive binaries that evolve more rapidly through LISA's band. Table~\ref{tab:multiband} gives both the expected counts and fractions.

In all four extrapolated models, binaries merging $10$--$30$ years after the LISA mission tend to have lower primary masses than those merging during it (Figure~\ref{fig:correlations}). Binaries merging during the mission and those merging $10$--$30$ years after its end are detected during the same four-year LISA observation. At a fixed observed frequency, binaries with smaller detector-frame chirp masses evolve more slowly \citep{Peters:1964Radiation}. At fixed primary mass, redshift, and observed frequency, a smaller companion lowers the chirp mass and lengthens the time to merger. The different companion-mass prescriptions in the primary-mass and symmetric models therefore affect the fraction of LISA detections that merge within each follow-up window.

We use these merger time cuts to identify potential multiband sources. We calculate their SNRs in one $40$-km CE detector, using a lower frequency cutoff of $5\,\mathrm{Hz}$ and requiring $\rho_{\rm CE}>8$ (Appendix~\ref{app:ce}). Assuming CE is operating at merger, more than $99.9\%$ of the GWTC-3 and GWTC-5.0 systems merging in each time window exceed this threshold, even at the lower end of the central $90\%$ credible interval on that fraction.

Among LISA detections that merge during the mission or within $30$ years after its end, the fraction detectable by CE is $95^{+5}_{-21}\%$ for \texttt{symBPL3P} extrapolated to $1000\msun$ and $99^{+1}_{-19}\%$ for \texttt{BPL2P} with a free high-mass slope. Some binaries in these extrapolations have such large redshifted masses that too little of their signal remains above $5\,\mathrm{Hz}$ to exceed the CE threshold. Table~\ref{tab:ce} gives the fractions for all models. For systems accessible in both bands, LISA's earlier observation can constrain the merger time and sky position before the merger \citep{Sesana:2016,Gerosa:2019Multiband,Klein:2022LastThreeYears}.

\section{Discussion}
\label{sec:discussion}

Using the released GWTC-3 and GWTC-5.0 population constraints, we forecast $\NThreeBaseEight$ and $\NOfficialBaseEight$ resolved binaries, respectively, in a four-year LISA mission at $\rho>\snrThr$. The updated population gives a higher median count, while retaining the expectation of a small catalog found in earlier GWTC-3 forecasts \citep{Buscicchio:2025SOBBH}. Binaries with primary masses between $100$ and $300\msun$ account for most of the increase in the posterior mean forecast, while the updated population below $100\msun$ also contributes (Section~\ref{sec:results:floor}).

LISA's sensitivity to the high-mass population depends on the companion masses as well as the primary masses. At fixed primary mass, redshift, and observed frequency, a heavier companion increases the chirp mass, giving a larger strain but a faster frequency evolution. Both affect the SNR accumulated during the four-year mission. In \texttt{symBPL3P}, the additional high-mass Gaussian enters the mass distribution used for both black holes. Even at comparable local merger rates, the four fits can therefore give different LISA counts because they assign different abundances to binaries with two massive components. The continuum, pairing, and low-mass cutoffs are also refitted, so this comparison does not measure the effect of the additional Gaussian alone. The fitted populations are restricted by the numerical precision of the likelihood: combinations that fail the variance requirement are excluded even if they are physically allowed. Its effect on the inferred population parameters and LISA counts has not been quantified (Appendix~\ref{app:rateinference}).

Extending the mass range does not by itself imply a large increase in the LISA count. With the merger-rate evolution held the same in the $300$ and $1000\msun$ calculations, the steep \texttt{BPL2P} primary-mass distribution adds $\NNtwoAddedEight$ expected detections above $300\msun$. The symmetric fits permit shallower high-mass component distributions and hence more binaries with massive companions. For \texttt{symBPL3P}, the expected count changes from $7.2^{+35}_{-3.2}$ to $\NSthreeExtEight$ under the same rate evolution (Table~\ref{tab:summary}). Its long upper interval comes from extrapolating populations that allow many more massive binaries than those giving counts near the median. These binaries add to the merger rate; the rates assigned to binaries with primary masses below $300\msun$ are unchanged.

Whether such a high-mass population exists depends on how these black holes form. Pair instability can prevent stellar collapse from producing black holes over a range of masses, while sufficiently massive progenitors can form black holes above that range \citep{Farmer:2019MindTheGap,HegerWoosley:2002,WoosleyHeger:2021Gap}. Remnants of earlier black hole mergers provide another route to high masses; their abundance depends in part on whether the remnants remain bound to their environment and merge again \citep{GerosaBerti:2019Escape,Gerosa:2021HierMerg}. Such repeated mergers are one possible interpretation of reported high-mass structure \citep{MaganaHernandez:2024Bump,Antonini:2025ClusterPopulation,Ginat:2026awh}. Our extrapolations do not impose a pair-instability gap or calculate the populations produced by these channels. The $300\msun$ boundary is the upper limit of the fits, not a physical gap boundary. The possible formation of black holes above the pair-instability gap motivates testing a different high-mass distribution, but does not prescribe its slope. Their contribution to LISA also depends on their merger rate and companion masses \citep{Gerosa:2021HierMerg,Antonini:2025ClusterPopulation}. The extrapolated counts also assume the fixed Madau--Fragos merger-rate evolution with redshift. Formation delays can change this evolution \citep{FishbachKalogera:2021TimeDelays}, and that uncertainty is not included in our intervals.

The O3 non-detection gives little additional information for the free-slope extrapolation considered here. Matching the rate density to \texttt{BPL2P} at $300\msun$ gives an expected O3 count of only $(5.9^{+72}_{-5.5})\times10^{-5}$ in the region covered by the nonspinning injections. The probability of zero detections is therefore close to unity, leaving the LISA forecast essentially unchanged. The wider LISA interval reflects the allowed mass distributions above the matching point. This result is specific to the adopted normalization and O3 sensitivity coverage. In particular, we note that the sensitivity used here does not extend to all masses and mass ratios in the LISA extrapolation (Appendix~\ref{app:o3}).

The largest \texttt{symBPL3P} forecasts also reach the limits of the signal and noise approximations. At large redshifted masses, the end of the inspiral moves to lower observed frequencies \citep{Husa:2016IMRPhenomD,Khan:2016IMRPhenomD}, whereas our LISA calculation uses a leading-order inspiral without a mass-dependent termination (Appendix~\ref{app:snr}). The total BBH background can also become comparable to the assumed noise. Adding this entire background to the noise reduces the count by about $13\%$ for a population near the $95$th percentile of expected count and $60\%$ for one near the $99$th percentile (Appendix~\ref{app:foreground}). The background used in these checks includes signals that could be subtracted, and the selected populations do not give a corrected credible interval. The extreme counts need to be assessed with mass-dependent signal termination and a calculation of detection and source subtraction together; the upper percentile is also only approximate because it is estimated from relatively few samples in the high-count tail (Appendix~\ref{app:sampling}).

For the released GWTC-5.0 population, $\OfficialMergerTen$ of LISA detections merge during the four-year mission or within ten years after its end. More than $99.9\%$ of those mergers exceed our reference CE threshold, even at the lower end of the central $90\%$ credible interval on that fraction, assuming CE is operating at merger. For this population, the number of joint observations is therefore limited mainly by which binaries LISA detects and when they merge, rather than by CE's sensitivity. The follow-up interval also selects different masses among LISA detections: in all four extrapolated models, binaries merging $10$--$30$ years after mission end tend to have lower primary masses than those merging during the mission (Figure~\ref{fig:correlations}). At a fixed observed frequency, smaller detector-frame chirp masses take longer to merge. Waiting longer for mergers therefore adds lower-mass systems to the potential multiband sample.

Even a small joint sample would follow the same binaries from their early inspiral to merger. LISA can provide information about the orbit before it has further circularized, while ground-based observations measure the late inspiral and remnant \citep{Breivik:2016FormationChannels, Nishizawa:2017FormationEccentricity,Buscicchio:2021LISABBHPE}. Combining the two bands can improve binary-parameter measurements and tests of general relativity \citep{Vitale:2016Multiband,Datta:2021MultibandGR}. Eccentricity, masses, and spins can help distinguish formation channels, although they need not uniquely identify a binary containing a previous merger remnant \citep{Gerosa:2021HierMerg}. Our circular-inspiral forecasts estimate the available sources, not the precision of those measurements. Further ground-based detections of systems such as GW190521 and GW231123 will constrain the abundance and companion masses of massive binaries, improving the forecasts for both LISA and joint observations.

\section{Conclusion}
\label{sec:conclusion}

We have used ground-based population constraints to forecast the number and masses of binary black holes detectable by LISA. For a \missionYr{}-year mission at $\rho>\snrThr$, we obtain $\NOfficialBaseEight$ expected detections using the baseline GWTC-5.0 constraints, compared with $\NThreeBaseEight$ using GWTC-3. The quoted errors enclose central $90\%$ intervals from population and merger-rate uncertainty.

We find that the assumed mass model can have a significant impact on the forecast. For our GWTC-5.0 fits, extrapolating the primary-mass models from $300$ to $1000\msun$ changes their median expected counts little, while widening their count intervals. The symmetric mass models allow a larger contribution from binaries with two massive components. The broadest forecasts arise when a weakly constrained high-mass distribution is extrapolated for both components. For the symmetric broken power law plus three peaks mass model, \texttt{symBPL3P}, the expected count changes from $7.2^{+35}_{-3.2}$ to $\NSthreeExtEight$, using the same Madau--Fragos merger-rate evolution in both cases and preserving the merger rate below $300\msun$. The broad upper interval reflects the uncertain high-mass abundance, but it is worth mentioning that our inspiral and noise approximations also limit the reliability of its extreme counts.

For the broken power law plus two peaks model, \texttt{BPL2P}, allowing the mass distribution above $300\msun$ to differ from its fitted form raises the upper end of the expected-count interval. When this extrapolation is matched to the fitted rate per unit primary mass at $300\msun$, it predicts negligible detections within the O3 search's sensitivity coverage. The O3 IMBH non-detection therefore leaves this forecast essentially unchanged.

For the released GWTC-5.0 population, $\OfficialMergerTen$ of LISA detections merge during the mission or within ten years after its end. Essentially all of these mergers would be detectable by our reference CE configuration if it is operating at merger. In all four extrapolated models we consider, binaries merging $10$--$30$ years after mission end tend to have lower primary masses than those merging during it. Further ground-based observations will constrain how often massive black holes merge and the masses of their companions, improving these forecasts. Observations in both frequency bands would follow the same binary from its early inspiral to merger, combining LISA's information about the orbit with ground-based measurements of the late inspiral and remnant.

\section*{Data Availability}

The \lisarates{} analysis code and data needed to reproduce the forecasts and figures will be made publicly available with this work.

\section*{Acknowledgments}

We thank the LISA Consortium and the LIGO--Virgo--KAGRA Collaboration for public data products. This work made use of publicly released GWTC-3 and GWTC-5.0 population posterior samples and the O3 IMBH search sensitivity data.

This material is based upon work supported by the National Aeronautics and Space Administration under Grant No. 22-LPS22-0025. Computations used the Vera cluster at the Pittsburgh Supercomputing Center, dedicated to the McWilliams Center for Cosmology and Astrophysics at Carnegie Mellon University. This work used Jetstream2 at Indiana University through allocation PHY240308 from the Advanced Cyberinfrastructure Coordination Ecosystem: Services \& Support (ACCESS) program, which is supported by National Science Foundation grants \#2138259, \#2138286, \#2138307, \#2137603, and \#2138296.

This research has made use of data or software obtained from the Gravitational Wave Open Science Center (\url{https://gwosc.org}), a service of the LIGO Scientific Collaboration, the Virgo Collaboration, and KAGRA \citep{Abbott:2023OpenData,LVK:2026O4bOpenData}. This material is based upon work supported by NSF's LIGO Laboratory which is a major facility fully funded by the National Science Foundation, as well as the Science and Technology Facilities Council (STFC) of the United Kingdom, the Max-Planck-Society (MPS), and the State of Niedersachsen/Germany for support of the construction of Advanced LIGO and construction and operation of the GEO600 detector. Additional support for Advanced LIGO was provided by the Australian Research Council. Virgo is funded, through the European Gravitational Observatory (EGO), by the French Centre National de Recherche Scientifique (CNRS), the Italian Istituto Nazionale di Fisica Nucleare (INFN) and the Dutch Nikhef, with contributions by institutions from Belgium, Germany, Greece, Hungary, Ireland, Japan, Monaco, Poland, Portugal, Spain. KAGRA is supported by Ministry of Education, Culture, Sports, Science and Technology (MEXT), Japan Society for the Promotion of Science (JSPS) in Japan; National Research Foundation (NRF) and Ministry of Science and ICT (MSIT) in Korea; Academia Sinica (AS) and National Science and Technology Council (NSTC) in Taiwan.

Claude Fable 5.1 (Anthropic) and Codex GPT-6 Astra (OpenAI) assisted with code development, numerical checks, figure preparation, and manuscript editing. The authors are responsible for the analysis and manuscript.

\facilities{LIGO, Virgo, KAGRA}

\software{\lisarates{} (this work),
\texttt{dynesty} \citep{Speagle:2020Dynesty}, \texttt{JAX} \citep{Bradbury:2018JAX}, \texttt{LEGWORK} \citep{Wagg:2022LEGWORK}, \texttt{LALSuite} \citep{LALSuite:2018,Wette:2020SWIGLAL}, \texttt{Astropy} \citep{Astropy:2022}, \texttt{NumPy} \citep{Harris:2020NumPy}, \texttt{SciPy} \citep{Virtanen:2020SciPy}, \texttt{Matplotlib} \citep{Hunter:2007Matplotlib}, \texttt{Seaborn} \citep{Waskom:2021Seaborn}}

\appendix

\section{Population Definitions and Merger Rates}
\label{app:popmodels}

We use source-frame component masses $m_1\geq m_2$ and mass ratio $q=m_2/m_1$. The population parameters, collectively denoted by $\Lambda$, specify the intrinsic mass and spin distributions and the redshift evolution of the merger rate. The density $p(m_1,q\mid\Lambda)$ describes all binaries in the model, whether or not they are detectable. For compactness, the equations we write show $p(m_1,q)$ in place of $p(m_1,q|\Lambda)$, but the mass distributions still depend on the fitted population parameters. The local merger rate $\mathcal R_0$ is the number of mergers per unit comoving volume and source-frame time at $z=0$, in $\mathrm{Gpc}^{-3}\,\mathrm{yr}^{-1}$. Thus $\mathcal R_0 p(m_1,q\mid\Lambda)\,dm_1\,dq$ gives the local rate in a small primary-mass and mass-ratio interval.

\subsection{Baseline Populations}

For GWTC-3, the primary-mass distribution is a mixture of a truncated power law and a Gaussian peak, multiplied by a low-mass taper \citep{TalbotThrane:2018Mass,Abbott:2021GWTC3pop}. The power-law slope, peak location and width, relative peak weight, mass cutoffs, and taper width are inferred parameters.

For a component mass $m$, let $x=m-m_{\min}$ be the distance above the lower mass cutoff. The taper rises from zero to one over a mass interval of width $\delta$ \citep{TalbotThrane:2018Mass}:
\begin{equation}
    \label{eq:smoothing}
    S(x;\delta)=
    \begin{cases}
    0, & x\leq0,\\
    \left[1+\exp\left(\dfrac{\delta}{x}
    +\dfrac{\delta}{x-\delta}\right)\right]^{-1},
    & 0<x<\delta,\\
    1, & x\geq\delta.
    \end{cases}
\end{equation}

For $\delta=0$, we use a sharp lower cutoff. The same taper is applied to the secondary mass $m_2=qm_1$. At fixed primary mass, the normalized mass-ratio density is \citep{TalbotThrane:2018Mass}
\begin{align}
    \label{eq:qpdf}
    p(q\mid m_1)&=\frac{q^{\beta_q}S(qm_1-m_{\min};\delta)}
    {Z_q(m_1)},\nonumber\\
    Z_q(m_1)&=\int_{m_{\min}/m_1}^{1}dq'\,
    q'^{\beta_q}S(q'm_1-m_{\min};\delta).
\end{align}

Its support is $m_{\min}/m_1\leq q\leq1$; it vanishes outside this interval. The exponent $\beta_q$ controls the relative abundance of unequal- and equal-mass binaries. The normalization $Z_q(m_1)$ depends on the primary mass because the allowed secondary-mass range changes with $m_1$.

The released GWTC-5.0 \textsc{Brokenpowerlaw+2Peaks} posterior uses a broken power law and two Gaussian peaks, with separate low-mass tapers for the primary and secondary masses \citep{Abbott:2025GWTC5pop}. We write the normalized broken power law as \citep{Abbott:2021GWTC2pop}
\begin{equation}
    \label{eq:brokenpl}
    b(m)=\frac{1}{Z_b}
    \begin{cases}
    (m/m_b)^{-\alpha_1},&m_{\rm low}\leq m<m_b,\\
    (m/m_b)^{-\alpha_2},&m_b\leq m\leq M_{\rm fit},\\
    0,&\text{otherwise}.
    \end{cases}
\end{equation}

Here $m_b$ is the break mass, and $\alpha_1$ and $\alpha_2$ set the slopes below and above it. The two power laws meet continuously at $m_b$. The lower edge is $m_{\rm low}$, the fitted upper limit is $M_{\rm fit}=300\msun$, and $Z_b$ is the integral of the piecewise shape over this interval.

We evaluate these distributions using the joint posterior samples released with each catalog, as described in Section~\ref{sec:popmodels}.

\subsection{Alternative Mass Models}
\label{app:fitinference}

The four new fits use the same broken power law, with either $K=2$ or $K=3$ Gaussian peaks. Before tapering, each Gaussian is normalized on the fitted mass interval:
\begin{equation}
    \mathcal N_{\rm t}(m;\mu_j,\sigma_j)=
    \frac{\exp[-(m-\mu_j)^2/(2\sigma_j^2)]}
    {\displaystyle\int_{m_{\rm low}}^{M_{\rm fit}}
    \exp[-(m'-\mu_j)^2/(2\sigma_j^2)]\,dm'}.
    \label{eq:truncatedpeak}
\end{equation}

This density is zero outside the integration limits. The parameters $\mu_j$ and $\sigma_j$ are the location and width of the underlying Gaussian, before truncation and tapering. The normalized mass distribution is
\begin{equation}
    B_K(m)=\frac{S(m-m_{\rm low};\delta_m)}{Z_K}
    \left[\lambda_0 b(m)+
    \sum_{j=1}^{K}\lambda_j
    \mathcal{N}_{\rm t}(m;\mu_j,\sigma_j)\right].
    \label{eq:componentmass}
\end{equation}

The coefficients $\lambda_j$ weight the component densities before the low-mass taper is applied: $\lambda_0$ weights the broken power law, $\lambda_j$ with $j\geq1$ weights peak $j$, and all weights are nonnegative with $\sum_{j=0}^{K}\lambda_j=1$. The normalization $Z_K$ is the integral of the numerator over $m_{\rm low}\leq m\leq M_{\rm fit}$, so that $\int B_K(m)\,dm=1$. The taper changes the relative integrals of the components; the $\lambda_j$ are therefore not generally the fractions remaining in each component after tapering.

For \texttt{BPL2P} and \texttt{BPL3P}, $B_K$ is the primary-mass density. Multiplying it by the companion's conditional mass-ratio density gives
\begin{equation}
    p_{\rm prim}(m_1,q)=B_K(m_1)\,p(q\mid m_1),
    \label{eq:nonsymmetric}
\end{equation}
where $p(q\mid m_1)$ follows Eq.~(\ref{eq:qpdf}), replacing $m_{\min}$ by the secondary-mass cutoff $m_{{\rm low},2}$ and $\delta$ by its taper width $\delta_{m,2}$. The secondary cutoff satisfies $3\msun\leq m_{{\rm low},2}\leq m_{\rm low}$.

For \texttt{symBPL2P} and \texttt{symBPL3P}, the same $B_K$ describes both component masses before pairing. We weight each pair by $q^\beta$. The resulting probability density in primary mass and mass ratio is
\begin{equation}
    p_{\rm comp}(m_1,q)=
    \frac{m_1 B_K(m_1)B_K(qm_1)q^\beta}{Z_{\rm comp}},
    \label{eq:symmetric}
\end{equation}

on the ordered domain $3\msun\leq m_2\leq m_1\leq300\msun$. Here $Z_{\rm comp}$ normalizes the distribution over ordered binary masses, so that $\int p_{\rm comp}(m_1,q)\,dm_1\,dq=1$ on the fitted mass range:
\begin{equation}
    Z_{\rm comp}=\int_{3\msun}^{M_{\rm fit}}dm_1\,B_K(m_1)
    \int_{3\msun}^{m_1}dm_2\,B_K(m_2)
    \left(\frac{m_2}{m_1}\right)^\beta.
    \label{eq:pairnorm}
\end{equation}
When extrapolating beyond $300\msun$, we keep this normalization fixed so that the rate below this mass is unchanged (Appendix~\ref{appendix:extend}).

The factor $m_1$ in Eq.~(\ref{eq:symmetric}) follows from $dm_2=m_1\,dq$ at fixed primary mass. Positive $\beta$ gives greater weight to pairs with similar masses, while negative $\beta$ favors unequal masses. In the symmetric models, $B_K$ describes both component masses before pairing. The primary-mass distribution also depends on the abundance of allowed companions and their pairing weights, as we obtain it by integrating $p_{\rm comp}(m_1,q)$ over $q$ at fixed $m_1$. It is therefore not simply $B_K(m_1)$. At fixed primary mass, the companion distribution is proportional to $B_K(qm_1)q^\beta$, so the peaks also affect which companions are favored.

These distinctions matter when comparing the peak fractions between models. We quote the fraction of primary black holes belonging to the third Gaussian in the intrinsic population, rather than its mixture weight. We calculate this fraction by integrating the third Gaussian's contribution to the normalized primary-mass distribution over the fitted mass range, including the low-mass taper in both models and pairing in the symmetric model. For the symmetric model, we use the third-Gaussian term of $B_K(m_1)$ in Eq.~(\ref{eq:symmetric}), but sum over all companion components through $B_K(qm_1)$ and keep the full population normalization $Z_{\rm comp}$. The companion need not belong to the third Gaussian.

\subsection{Population Inference and the Local Merger Rate}
\label{app:rateinference}

All four fits use the same 259 GWTC-5.0 events and the cumulative O1--O4b injection set \citep{LVK:2026GWTC5Sensitivity}, which contains simulated signals used to estimate the probability of detection by the ground-based searches. We infer the mass, spin, and redshift parameters with a hierarchical likelihood that integrates over the local merger rate using the prior $p(\mathcal R_0)\propto1/\mathcal R_0$ and accounts for selection \citep{Mandel:2019Selection,ThraneTalbot:2019Inference,Farr:2019Selection}. The effective spin is $\chi_{\rm eff}=(\chi_1+q\chi_2)/(1+q)$ \citep{Racine:2008,Ajith:2011}, where $\chi_1$ and $\chi_2$ are the dimensionless component spins projected along the orbital angular momentum. Its density is a Gaussian with location $\mu_\chi$ and width $\sigma_\chi$, normalized on $[-1,1]$ \citep{RouletZaldarriaga:2019,Miller:2020LowSpin}. Within each population, this spin distribution is independent of mass and redshift, and the mass distribution does not evolve with redshift. The merger-rate density follows $\mathcal R(z)=\mathcal R_0(1+z)^\kappa$ over the fitted range $0\leq z\leq1.9$; $\kappa$ controls its growth or decline. The fits use nested sampling \citep{Skilling:2006} with \texttt{dynesty} \citep{Speagle:2020Dynesty}. Table~\ref{tab:fitpriors} lists the mass, spin, and redshift priors.

\begin{deluxetable}{lc}[t]
\tablecaption{Priors for the four GWTC-5.0 fits. $U[a,b]$ denotes a uniform density.\label{tab:fitpriors}}
\tablehead{\colhead{Parameter} & \colhead{Prior}}
\startdata
Power-law slopes $\alpha_1,\alpha_2$ & $U[-4,12]$ \\
Break mass $m_b/M_\odot$ & $U[20,50]$ \\
Primary or common lower edge $m_{\rm low}/M_\odot$ & $U[3,10]$ \\
Taper widths $\delta_m/M_\odot$ & $U[0,10]$ \\
Secondary edge $m_{{\rm low},2}/M_\odot$ & $U[3,m_{\rm low}]$ \\
First peak mean $\mu_1/M_\odot$ & $U[5,20]$ \\
Second peak mean $\mu_2/M_\odot$ & $U[25,60]$ \\
Third peak mean $\mu_3/M_\odot$ & $U[45,90]$ \\
Peak widths $\sigma_j/M_\odot$ & $U[0,10]$ \\
Mixture weights $(\lambda_0,\ldots,\lambda_K)$ & Dirichlet$(1,\ldots,1)$ \\
Mass-ratio or pairing exponent $\beta_q$ or $\beta$ & $U[-2,7]$ \\
Spin mean $\mu_\chi$ & $U[-1,1]$ \\
Log spin width $\ln\sigma_\chi$ & $U[-5,0]$ \\
Rate exponent $\kappa$ & $U[-10,10]$ \\
\enddata
\tablecomments{The primary-mass models have separate lower mass cutoffs and smoothing widths for the primary and secondary. The symmetric models use the same mass distribution, cutoff, and smoothing width for both black holes. The Dirichlet prior is uniform over nonnegative mixture weights that sum to one. The third-peak parameters apply only to the three-peak models. All fits have a fixed upper mass limit of $300M_\odot$.}
\end{deluxetable}

\newpage

The hierarchical likelihood is computed by reweighting single event-posterior samples and simulated signals. Let $I_i$ denote the population-weighted likelihood integral for event $i$ and $\xi$ the population-averaged detection efficiency. After marginalizing over the merger rate, the log likelihood contains $\sum_i\ln I_i-N_{\rm obs}\ln\xi$, where $N_{\rm obs}=259$ is the number of events. Assuming independent Monte Carlo estimates, we estimate the numerical variance as \citep{TalbotGolomb:2023Likelihood}
\begin{equation}
    \widehat\sigma^2_{\ln\mathcal L}\simeq
    \sum_i\widehat{\mathrm{Var}}(\ln I_i)
    +N_{\rm obs}^2\frac{\widehat{\mathrm{Var}}(\xi)}{\xi^2}.
    \label{eq:likelihoodvariance}
\end{equation}
During inference, we set the likelihood to zero wherever this estimate is at least one. This avoids using poorly determined likelihood values, but also restricts the populations the fit can explore \citep{Abbott:2025GWTC4pop}.

Many posterior samples have estimated log-likelihood variance close to the imposed limit, $\widehat\sigma^2_{\ln\mathcal L}=1$. The forecast intervals are conditional on this requirement; we have not quantified how it changes the inferred populations or LISA counts. The variance requirement controls Monte Carlo error in the standard hierarchical likelihood, arising from both the selection-function injections and the event-level parameter-estimation samples. Following \citet{TalbotGolomb:2023Likelihood} and \citet{EssickFarr:2022Precision}, these sample sets must provide sufficiently precise likelihood estimates to avoid biasing population constraints. The imposed limit of 1 on the estimated log-likelihood variance is a numerical reliability criterion, not a physical population constraint, and it does not by itself establish that the posterior is unbiased \citep{HeinzelVitale:2025MonteCarlo}. Testing populations excluded by this requirement requires more accurate likelihood estimates, rather than simply relaxing the threshold.

We reconstruct a local merger-rate distribution for each draw of the population parameters. A population that is harder to detect requires a larger intrinsic rate to account for the same observed catalog. This dependence is described by the sensitive spacetime volume \citep{Tiwari:2018VT,Farr:2019Selection}
\begin{equation}
    \langle VT\rangle_\Lambda={T_{\rm GW}}
    \int_0^{1.9}dz\,\frac{dV_c}{dz}(1+z)^{\kappa-1}
    \overline P_{\rm det}(z\mid\Lambda).
    \label{eq:sensitivevolume}
\end{equation}
Here $T_{\rm GW}$ is the ground-based observing duration and $dV_c/dz$ is the full-sky comoving volume per unit redshift \citep{Hogg:1999}. The detection probability $\overline P_{\rm det}$ is averaged over the population's masses and spins, source orientation, and the observing periods. The factor $(1+z)^\kappa$ describes the rate evolution, while $(1+z)^{-1}$ converts source-frame time to observer time. We estimate $\langle VT\rangle_\Lambda$, in $\mathrm{Gpc}^3\,\mathrm{yr}$, by weighting the recovered injections by the population density relative to the injection density, normalizing by the total number generated, and including the observing duration \citep{Tiwari:2018VT,Essick:2025O4Injections}.

For fixed $\Lambda$, the expected number of ground-based detections is $\mathcal R_0\langle VT\rangle_\Lambda$. The Poisson count likelihood and the prior $p(\mathcal R_0)\propto1/\mathcal R_0$ give the conditional rate posterior \citep{Mandel:2019Selection}
\begin{equation}
    p(\mathcal R_0\mid\Lambda,\{d_i\})
    =\frac{\langle VT\rangle_\Lambda^{N_{\rm obs}}}
    {\Gamma(N_{\rm obs})}\,
    \mathcal R_0^{N_{\rm obs}-1}
    e^{-\mathcal R_0\langle VT\rangle_\Lambda}.
    \label{eq:rategamma}
\end{equation}

Here $\{d_i\}$ denotes the event data, $N_{\rm obs}=259$ is the number of events, and $\Gamma$ is the Gamma function. This is a Gamma density on $\mathcal R_0>0$, with shape $N_{\rm obs}$, rate parameter $\langle VT\rangle_\Lambda$, and conditional mean $N_{\rm obs}/\langle VT\rangle_\Lambda$. Both the expected-count intervals and the shaded mass-distribution bands in the Figures include the conditional rate uncertainty in Eq.~(\ref{eq:rategamma}). We integrate over this distribution for the count intervals. For the bands in Figure~\ref{fig:massmodels}, we pair each population sample with one rate drawn from its own conditional distribution and calculate the spread of the resulting merger-rate and detected-count densities.

\subsection{Extrapolated Merger Rates and Redshift Evolution}
\label{appendix:extend}

Let $\widetilde p(m_1,q\mid\Lambda)$ denote the mass density obtained by extending the upper mass limits of the fitted functions to $1000\msun$, retaining the original normalization of each mixture component and of $B_K$. Below $300\msun$, $\widetilde p$ equals the fitted density exactly. Above this boundary, we retain the fitted slopes, Gaussian parameters, and pairing prescription.

For \texttt{BPL2P} and \texttt{BPL3P}, the conditional mass-ratio density is normalized over the allowed companion masses for each primary mass, including primaries above $300\msun$. For \texttt{symBPL2P} and \texttt{symBPL3P}, we use the value of $Z_{\rm comp}$ calculated with both masses below $300\msun$. Recalculating it over the enlarged mass range would reduce the rate of the original binaries, contrary to the extrapolation defined here. The extrapolated local rate is
\begin{equation}
    \frac{d^2\mathcal R_{\rm ext}}{dm_1\,dq}
    =\mathcal R_0\,\widetilde p(m_1,q\mid\Lambda),
    \qquad m_1\leq1000\msun.
    \label{eq:additive_rate}
\end{equation}

Unlike the fitted density, $\widetilde p$ is not normalized to one over the enlarged mass range. The total local rate becomes
\begin{align}
    \mathcal R_{\rm tot,ext}&=\mathcal R_0+\Delta\mathcal R,\nonumber\\
    \Delta\mathcal R&=\mathcal R_0
    \int_{300\msun}^{1000\msun}dm_1\int_0^1dq\,
    \widetilde p(m_1,q\mid\Lambda).
    \label{eq:addedrate}
\end{align}
Thus $\mathcal R_0$ still denotes the rate on the original mass range, and $\Delta\mathcal R$ is the rate of binaries with a primary above $300\msun$. The rate for every pair with $m_1\leq300\msun$ is unchanged, as is the rate per unit primary mass below this boundary. In all four models, an added binary may have one or both components above $300\msun$. The abundance of secondaries below this boundary can therefore increase when they pair with newly added primaries.

For all extrapolated populations and their matched $300\msun$ comparisons, we obtain the rate at redshift $z$ by multiplying the local mass-dependent rate by $g(z)$, where
\begin{align}
    g(z)&=\frac{\psi(z)}{\psi(0)},\\
    \psi(z)&=\frac{(1+z)^d}
    {1+\dfrac{d}{|r|}
    \left(\dfrac{1+z}{1+z_p}\right)^{d-r}},
    \quad (d,r,z_p)=(2.6,-3.6,2.04).
    \label{eq:mf17}
\end{align}

Here $\psi(z)$ is a dimensionless function with the star-formation-history shape of \citet{Madau:2017SFH}, parameterized by its peak redshift as in \citet{Babak:2023SGWB}. Only the ratio $\psi(z)/\psi(0)$ enters the merger rate so no absolute star-formation-rate normalization is required. The exponents $d$ and $r$ give the low- and high-redshift slopes in $1+z$, and $z_p$ is the redshift of the maximum. These parameters are fixed, not inferred from the LVK catalog. The $300$ and $1000\msun$ forecasts use the same $g(z)$, so their difference is due to the contribution of the added masses. We note that this assumed merger-rate history does not include a distribution of delays between formation and merger, and the forecast intervals do not include uncertainty in that history.

\section{LISA Noise and Signal-to-Noise Ratio}
\label{app:noise}

\subsection{Instrumental Noise and Galactic Foreground}

We use the equal-arm effective strain sensitivity in Eq.~(56) of \citet{Babak:2021SNR}, with arm length $L=2.5\times10^9\,$m, acceleration-noise level $3\times10^{-15}\,\mathrm{m\,s^{-2}\,Hz^{-1/2}}$, and optical-metrology noise level $15\times10^{-12}\,\mathrm{m\,Hz^{-1/2}}$. The effective strain-noise spectrum $S_{n,\rm inst}$ includes the sky- and polarization-averaged response for one channel, without the improvement from combining two channels. We add the four-year foreground from unresolved Galactic double white dwarfs of \citet{Robson:2019LISASens}, using the single-channel amplitude $A=1.80\times10^{-44}\,\mathrm{Hz}^{-1}$ of \citet{Cornish:2017GalacticBinary}. The Galactic foreground is expressed in the same single-channel convention. Writing its spectrum as $S_c$, we use $S_n=S_{n,\rm inst}+S_c$. The factor of two in squared SNR for two effective channels enters in Eq.~(\ref{eq:snriota}), $\rho^2=2\times4\,\Xi^2\rho_0^2$. No additional channel factor is applied to $S_n$ or $\rho$.

\begin{figure*}[t]
\centering
\includegraphics[width=\textwidth]{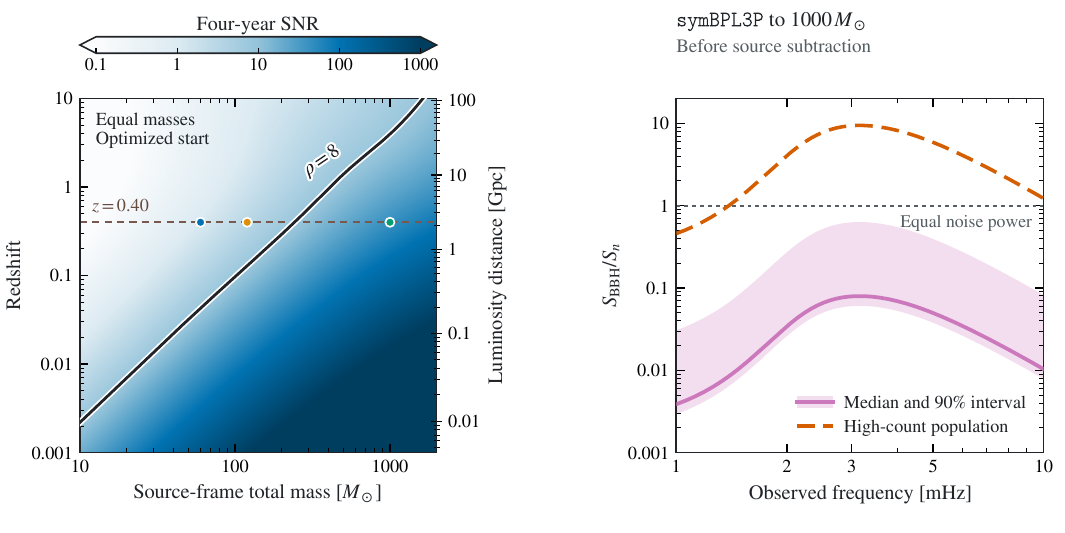}
\caption{LISA detection reach and the binary black hole (BBH) background. \emph{Left:} four-year SNR for equal-mass BBHs, maximizing over starting frequency and averaging the response as described in Appendix~\ref{app:snr}. The contour marks $\rho=8$; points at $z=0.40$ locate the equal-mass examples in Figure~\ref{fig:lisasensitivity}. \emph{Right:} total BBH background power spectral density (PSD) divided by the combined PSD of instrumental noise and unresolved Galactic double white dwarfs for \texttt{symBPL3P} extrapolated to $1000\msun$. The solid line and band give the median and central $90\%$ population interval; the dashed curve shows a population near the $99$th percentile of expected count. Signals from individually detectable BBHs have not been subtracted.}
\label{fig:noisehorizon}
\end{figure*}

\subsection{Signal-to-Noise Ratio}
\label{app:snr}

For source-frame chirp mass $\Mc$ and source-frame time to coalescence $\tau$, the leading-order circular inspiral gives the source-frame frequency \citep{Peters:1964Radiation,CutlerFlanagan:1994Parameters}
\begin{equation}
    f(\tau)=\left[\frac{256}{5}\pi^{8/3}
    (G\Mc/c^3)^{5/3}\tau\right]^{-3/8}.
    \label{eq:ftau}
\end{equation}

A source at redshift $z$ evolves for $T_{\rm obs}/(1+z)$ in its own frame during the mission, or until merger if that occurs first. We convert to observed frequency using $f_{\rm det}=f/(1+z)$ and integrate over the frequencies traversed during the mission, restricted to the adopted LISA band, $3\times10^{-5}$--$0.5$ Hz. We use the leading-order inspiral waveform up to the upper frequency of this band without a mass-dependent termination. In the integral below, $f$ denotes the observed frequency. We define
\begin{equation}
    I=\int_{f_{\rm in,det}}^{f_{\rm out,det}}
    \frac{f^{-7/3}}{S_n(f)}\,df.
    \label{eq:Iintegral}
\end{equation}

Here the limits are the observed frequencies at the start and end of the observation, restricted to the adopted band. The dimensionless SNR factor is
\begin{equation}
    \rho_0=\frac{2}{\pi^{2/3}}\sqrt{\frac{5}{96}}\,
    \frac{(G\Mc_{\rm det})^{5/6}}{d_Lc^{3/2}}\sqrt I,
    \qquad \Mc_{\rm det}=(1+z)\Mc.
    \label{eq:amp}
\end{equation}

At inclination $\iota$,
\begin{equation}
    \rho=2\sqrt2\,\Xi(\iota)\rho_0,\qquad
    \Xi(\iota)=\sqrt{\left(\frac{1+\cos^2\iota}{2}\right)^2
    +\cos^2\iota}.
    \label{eq:snriota}
\end{equation}

These expressions average over sky position and polarization and include both effective channels \citep{Babak:2021SNR}. Isotropic inclinations give $\langle\Xi^2\rangle=4/5$ and $\langle\rho^2\rangle_\iota=(32/5)\rho_0^2$, while the face-on value is $\rho^2=16\rho_0^2$. No additional channel factor is applied.

For matched source parameters, single-channel noise curve, and frequency limits, our face-on $\rho^2$ is a factor of two below the approximate face-on $\rho^2$ in Eq.~(2.2) of \citet{Buscicchio:2025SOBBH}. They use that expression for preliminary catalog selection and recompute SNRs from simulated instrument data for their reported detection forecasts. As an independent check of the SNR normalization, we compare with \texttt{LEGWORK}, a code for gravitational waves from stellar-origin binaries \citep{Wagg:2022LEGWORK}. For circular binaries whose frequency changes little during the mission, the SNRs agree to better than $1\%$ when the source parameters and noise curves are matched.

For the four reference binaries in Figure~\ref{fig:lisasensitivity}, we use effective characteristic strain to display the same SNR calculation. Writing Eq.~(\ref{eq:amp}) as $\rho_0=\mathcal A\sqrt I$, we define \begin{equation}
    {h_n(f)=\sqrt{fS_n(f)},\qquad
    h_{\rm eff}(f)=\sqrt{\frac{32}{5}}\,\mathcal A f^{-1/6}.}
    \label{eq:heff}
\end{equation}
The ratio of these curves gives the SNR accumulated during the observation: \begin{equation}
    {\langle\rho^2\rangle_\iota=
    \int_{f_{\rm in,det}}^{f_{\rm out,det}}
    \left[\frac{h_{\rm eff}(f)}{h_n(f)}\right]^2 d\ln f.}
    \label{eq:heffsnr}
\end{equation}
The reference curves and bands include both channels and the root-mean-square response over isotropic inclinations. For each fixed source-frame mass pair, the bold curve uses $z=0.40$ and the band spans $0.1\leq z\leq1.0$. These bands compare redshifts; they are not uncertainty intervals. We recalculate the luminosity distance, detector-frame chirp mass, and starting frequency at each redshift, keeping the observation in the final four observer years before merger.

For the purple sample, we choose one \texttt{BPL2P} population near the posterior medians of its mass-distribution parameters. We extrapolate it to $1000\msun$ with the Madau--Fragos merger-rate history. We sample masses, redshifts, merger times, and isotropic inclinations, weighting sources by their expected abundance during the mission. Each track covers the frequencies traversed in four observer years, or until the signal cutoff if reached earlier; merger need not occur during the mission. We consider sources with total four-year SNR $0.1\leq\rho\leq8$ and a track intersecting the displayed frequency range. The lower cut selects the louder part of the undetected population for illustration. The shading shows the smoothed, population-weighted distribution of individual tracks in frequency and strain in the upper panel, and in frequency and accumulated SNR in the lower panel. Each panel is scaled to its own maximum for display. The lower-panel SNR is accumulated from the start of each source's observation. No BBH contribution is added to the noise in Figure~\ref{fig:lisasensitivity}, but the summed BBH background is treated separately in Appendix~\ref{app:foreground}.

Figure~\ref{fig:noisehorizon} (left) instead chooses the starting frequency that maximizes the four-year SNR for each mass and redshift. Its $\rho=8$ contour is therefore a reach estimate with the averaged response, whereas the four reference binaries in Figure~\ref{fig:lisasensitivity} are observed during their final four years before merger. The population forecasts integrate over merger times and inclinations. The horizon search extends to $z=14$; the plot stops at $z=10$. This numerical search limit is separate from the redshift integration used for the population forecasts.

\subsection{Binary Black Hole Background}
\label{app:foreground}

To check whether the binary black hole (BBH) background could affect our largest detection forecasts, we calculate the total circular-inspiral background for \texttt{symBPL3P} extrapolated to $1000\msun$, with the adopted merger-rate evolution with redshift, using the relation between merger rate and emitted gravitational-wave energy \citep{Phinney:2001Background,Babak:2023SGWB}. This background includes signals from individually detectable BBHs. Subtracting their signals would reduce the remaining background, but we do not estimate the residual background here.

Figure~\ref{fig:noisehorizon} (right) shows the BBH background power spectral density (PSD), $S_{\rm BBH}$, relative to $S_n$, the sum of instrumental noise and the foreground from unresolved Galactic double white dwarfs. The band includes uncertainty in the fitted mass distribution and merger rate, conditional on the extrapolation and fixed merger-rate evolution with redshift. At $3$ mHz, $S_{\rm BBH}/S_n=0.08^{+0.55}_{-0.02}$ (median and central $90\%$ interval). The dashed curve is a separate example selected near the $99$th percentile of expected detection count; it is not the upper edge of the background interval.

To test the effect on individual BBH detections, we repeat the SNR calculation for populations near the median, $95$th, and $99$th percentiles of expected count, after adding each population's total BBH background to the assumed noise. This treats the combined BBH signal as stationary Gaussian noise, including signals from individually detectable BBHs. The expected counts above $\rho=8$ decrease by about $5\%$, $13\%$, and $60\%$, respectively. These checks evaluate individual populations, not posterior credible intervals or the residual noise after source subtraction. The most extreme forecasts therefore require a calculation of detection and source subtraction together \citep{Babak:2023SGWB}.

\section{Expected Counts and Population Uncertainty}
\label{app:sampling}

Let $r(m_1,q,z)$ be the merger rate per unit source-frame time, comoving volume, primary mass, and mass ratio. For each population and rate draw, the expected number observable by LISA is
\begin{equation}
    \mu=\int dm_1\,dq\,dz\,dt_{\rm c,det}\,
    \frac{r(m_1,q,z)}{1+z}\frac{dV_c}{dz}
    P_\iota(\rho>\rho_{\rm thr}).
    \label{eq:countintegral}
\end{equation}
The factor $P_\iota$ is the fraction of isotropic inclinations for  which Eq.~(\ref{eq:snriota}) exceeds the SNR threshold, using the frequencies traversed during the mission. The time to coalescence $t_{\rm c,det}$ is measured in the observer frame from mission start. We use the merger rate at redshift $z$ to infer the number of binaries that are still inspiraling at that redshift. We assume that this rate changes negligibly over their remaining time to merger, which can extend beyond the fiducial four-year mission \citep{Babak:2023SGWB,Buscicchio:2025SOBBH}. At fixed masses and redshift, equal intervals of time to coalescence therefore contain the same expected number of BBHs before applying the LISA selection. We assume the redshift dependence of $r(m_1,q,z)$, allowing the merger rate to vary across cosmic time.

We evaluate the mass, redshift, and time integrals numerically. For the catalog forecasts and fitted populations, we integrate the inferred power-law rate evolution over $0\leq z\leq1.9$. For the extrapolations and their matched $300\msun$ comparisons, we use Eq.~(\ref{eq:mf17}) and include redshifts and merger times throughout the detectable range. The upper time-to-merger bound follows from the low-frequency edge of the LISA band. We choose the redshift bound so that the maximum possible SNR is below threshold beyond it. The mission limits the observed inspiral to at most four years, but we stop earlier if the binary merges or leaves the LISA frequency range used in the calculation.

The final baseline forecasts use all 11,184 released GWTC-3 samples and all 8,200 released GWTC-5.0 samples. The alternative forecasts use 4,000 population samples per model, with conditional rate uncertainty from Eq.~(\ref{eq:rategamma}). We assess numerical stability with finer quadrature, and sensitivity to the finite population sample by comparing subsets and bootstrap resamples of the stored draws. For \texttt{symBPL3P} extrapolated to $1000\msun$, the upper end of the central $90\%$ interval on the expected count is about $3500$ detections. Resampling the inferred populations with replacement gives a central $90\%$ range of approximately $3000$--$4200$ for this endpoint. This measures uncertainty in the endpoint caused by the finite population sample, not an additional interval on the expected count. We therefore treat the quoted upper endpoint as approximate.

For a time $\Delta t$ after mission end, the merger-window count uses $0\leq t_{\rm c,det}\leq T_{\rm obs}+\Delta t$ in Eq.~(\ref{eq:countintegral}). For each population, the expected fraction merging in that window is $f_{\Delta t}=\mu_{\Delta t}/\mu_{\rm all}$, where $\mu_{\Delta t}$ and $\mu_{\rm all}$ are calculated for the same population parameters. The overall rate normalization cancels in this ratio. We report the median and central $90\%$ interval of $f_{\Delta t}$ across the population posterior.

The quoted $90\%$ count intervals describe uncertainty in $\mu$ from the inferred population parameters and merger rate. Neither these intervals nor those on $f_{\Delta t}$ include Poisson fluctuations in a realized LISA catalog. To calculate the probability of an empty LISA binary black hole catalog, we include these fluctuations by averaging $P(N=0\mid\mu)=e^{-\mu}$ over the population and rate posterior:
\begin{equation}
P(N=0)=\int d\Lambda\,d\mathcal R_0\,
p(\Lambda,\mathcal R_0\mid\{d_i\})e^{-\mu(\Lambda,\mathcal R_0)}.
\label{eq:predictive}
\end{equation}

\section{Cosmic Explorer Detectability}
\label{app:ce}

We use the published strain sensitivity for a $40$-km CE detector (CE-T2000017-v9; \citealt{Srivastava:2022CETunable})\footnote{\url{https://dcc.cosmicexplorer.org/public/0163/T2000017/009/ce_sensitivity.zip}}. The adopted sensitivity curve is evaluated at $15^\circ$ from normal incidence. We use \textsc{IMRPhenomXHM} waveforms, which include inspiral, merger, ringdown, and higher harmonics \citep{GarciaQuiros:2020IMRPhenomXHM}. We calculate the CE SNR through:
\begin{equation}
 \rho_{\rm CE}^2 =
 4\int_{5\,{\rm Hz}}^{2048\,{\rm Hz}}
 \frac{|F_+\widetilde h_+(f)+F_\times\widetilde h_\times(f)|^2}
 {S_{n,\rm CE}(f)}\,df ,
 \label{eq:ce_snr}
\end{equation}
where $F_+$ and $F_\times$ are the detector antenna factors and $S_{n,\rm CE}$ is the one-sided strain-noise PSD \citep{CutlerFlanagan:1994Parameters}. The waveform uses the same luminosity distance as the LISA calculation. We use the long-wavelength antenna response of an L-shaped detector with the published strain curve and do not include the full direction-dependent finite-arm response.

For the four fitted models and the free-slope extrapolation, we set the aligned component spins to $\chi_{1z}=\chi_{2z}=\chi_{\rm eff}$, with $\chi_{\rm eff}$ drawn from the fitted distribution. For the released catalog populations, we use the aligned projections of their inferred spin magnitudes and tilts. Our prescriptions neglect precession. We integrate the CE detection probability over isotropic sky positions and polarization angles, assuming the inclination distribution selected by LISA.

Let $\mu_{\rm LISA+CE,\Delta t}$ be the expected count after imposing $\rho_{\rm CE}>8$ in addition to the LISA selection and merger-time window in Eq.~(\ref{eq:countintegral}). Table~\ref{tab:ce} reports $F_{\rm CE}=\mu_{\rm LISA+CE,\Delta t}/\mu_{\Delta t}$: the fraction detectable by CE among LISA sources that merge in that window.

For the released GWTC-5.0 population, the median CE SNR among LISA sources merging during the mission or within $30$ years after its end is $(2.5^{+0.9}_{-1.1})\times10^3$ (with the spread indicating the population uncertainty on this median).  For mergers during the mission or within ten years after its end, the fraction of all LISA detections that also exceed the CE threshold is $39^{+22}_{-11}\%$, unchanged at the quoted precision from the fraction without a CE requirement. The lower ends of the central $90\%$ credible intervals on the CE detection fractions also exceed $99.9\%$ for all four fits restricted to $300\msun$, with either rate history.

These estimates assume that CE is operating when the binaries merge. The intervals include population uncertainty, conditional on the adopted sensitivity and spin prescription.

\begin{table*}[t]
\centering
\caption{Fraction detectable by the reference $40$-km CE among binaries detected by LISA at $\rho>8$ and merging in different time windows with respect to the LISA mission timeline.  Values are medians and central $90\%$ population intervals.  The released reported GWTC-3 and GWTC-5.0 populations retain their original mass limits and rate histories, whereas the other rows use extrapolations to $1000\msun$ with the adopted Madau--Fragos rate history. }
\label{tab:ce}
{\renewcommand{\arraystretch}{1.3}\begin{tabular}{lccc}
\hline
Population & During mission & Mission + 10 yr & Mission + 30 yr \\
\hline
GWTC-3 & $>99.9\%$ & $>99.9\%$ & $>99.9\%$ \\
\textcolor{black}{GWTC-5.0 (LVK)} & $>99.9\%$ & $>99.9\%$ & $>99.9\%$ \\
\texttt{BPL2P} & $99.3^{+0.7}_{-3.3}\%$ & $99.7^{+0.3}_{-2.3}\%$ & $99.8^{+0.2}_{-2.1}\%$ \\
\texttt{BPL3P} & $98.7^{+1.3}_{-7.6}\%$ & $99.4^{+0.6}_{-6.2}\%$ & $99.6^{+0.4}_{-5.9}\%$ \\
\texttt{symBPL2P} & $97.5^{+2.4}_{-8.1}\%$ & $98.7^{+1.3}_{-7.1}\%$ & $98.9^{+1.1}_{-7.1}\%$ \\
\texttt{symBPL3P} & $92^{+8}_{-21}\%$ & $95^{+5}_{-20}\%$ & $95^{+5}_{-21}\%$ \\
\textcolor{black}{\texttt{BPL2P} (free slope)} & $96^{+4}_{-21}\%$ & $98^{+2}_{-19}\%$ & $99^{+1}_{-19}\%$ \\
\hline
\end{tabular}}
\end{table*}

\section{Free-Slope Extrapolation and IMBH Limits}
\label{app:o3}

For the free-slope extrapolation of \texttt{BPL2P}, let $\mathcal R_{\rm ref}$ denote its fitted rate. The primary-mass rate above $m_{\rm j}=300\msun$ is
\begin{equation}
\frac{d\mathcal R_{\rm tail}}{dm_1}
=\left.\frac{d\mathcal R_{\rm ref}}{dm_1}\right|_{m_{\rm j}}
\left(\frac{m_1}{m_{\rm j}}\right)^{-\alpha},
\qquad m_{\rm j}<m_1\leq1000\msun.
\label{eq:tail}
\end{equation}

We retain the reference model's conditional mass-ratio prescription and use Eq.~(\ref{eq:mf17}) for rate evolution. The original population below $m_{\rm j}$ is unchanged. For each draw of the reference model's shape and rate, we start with $\alpha\sim U[-2,8]$ and update it by
\begin{equation}
p(\alpha\mid\Lambda,\mathcal R_0,0_{\rm O3})
\propto p(\alpha)\exp[-\mu_{\rm O3}
(\alpha,\Lambda,\mathcal R_0)].
\label{eq:tailupdate}
\end{equation}

Here $0_{\rm O3}$ denotes zero detections with $m_1>300\msun$, and $\mu_{\rm O3}$ is the expected number of O3 detections from this added population. For a given fitted population and merger rate, we normalize the distribution above over $\alpha$. We average these slope distributions over the posterior from our \texttt{BPL2P} fit to GWTC-5.0, keeping its population and rate sample weights unchanged. This update uses the O3 non-detection to reweight only the extrapolation slope.

GW190521 remains part of the fitted population; the zero count applies only to binaries with $m_1>300\msun$ \citep{Abbott:2020GW190521,Abbott:2022IMBHO3}. We calculate $\mu_{\rm O3}$ by integrating the tail rate against the nonspinning sensitive spacetime volume from the released O3 IMBH injections \citep{Abbott:2022IMBHO3}.\footnote{LIGO-P2100179-v6, \texttt{O3IMBHSensitivity.tar.gz}: \url{https://dcc.ligo.org/LIGO-P2100179/public/}} We weight the recovered injections by the redshift dependence in Eq.~(\ref{eq:mf17}), normalized to its local value, retaining the release's spacetime-volume normalization and cosmology. The published sensitivity nodes and their upper limits provide a check of this reconstruction. They are limits for specified source populations, not upper bounds on $d\mathcal R/dm_1$ at every mass.

We interpolate the sensitive spacetime volume only within the region enclosed by the sampled mass and mass-ratio combinations in $(\log m_1,\log q)$. Its direct coverage ends at $m_1=400\msun$ and has $q\geq0.1$, with additional restrictions from the node layout. No direct O3 sensitivity is assigned outside this region, because the grid does not supply it. The calculation is conditional on the nonspinning exposure and does not marginalize over the spin dependence of the search. At the rate densities inferred at $300\msun$, $\mu_{\rm O3}\ll1$ across the tested slopes. The O3 non-detection therefore has a negligible effect on the LISA forecast.

\clearpage
\section{Forecast Tables}
\label{app:tables}

\startlongtable
\begin{deluxetable}{lrlcc}
\tablewidth{\textwidth}
\tablecaption{Expected LISA detections in four years.
Entries give medians with lower and upper errors enclosing the central
$90\%$ credible interval.
\label{tab:summary}}
\tablehead{\colhead{Population} & \colhead{$m_{1,\max}/M_\odot$} & \colhead{Rate evolution}
& \colhead{$\rho>8$} & \colhead{$\rho>15$}}
\startdata
GWTC-3 & Varies & Fitted power law & $4.3^{+3.6}_{-2.0}$ & $0.63^{+0.54}_{-0.29}$ \\
GWTC-5.0 (LVK) & $300$ & Fitted power law & $6.6^{+5.5}_{-2.2}$ & $0.91^{+0.64}_{-0.30}$ \\
\texttt{BPL2P} & $300$ & Fitted power law & $5.7^{+4.9}_{-1.9}$ & $0.79^{+0.57}_{-0.25}$ \\
\texttt{BPL2P} & $300$ & Madau--Fragos & $5.6^{+5.3}_{-1.9}$ & $0.78^{+0.59}_{-0.26}$ \\
\texttt{BPL2P} & $1000$ & Madau--Fragos & $5.9^{+9.5}_{-2.2}$ & $0.84^{+1.5}_{-0.3}$ \\
\texttt{BPL3P} & $300$ & Fitted power law & $5.9^{+6.6}_{-1.9}$ & $0.81^{+0.72}_{-0.26}$ \\
\texttt{BPL3P} & $300$ & Madau--Fragos & $5.8^{+7}_{-2}$ & $0.8^{+0.74}_{-0.27}$ \\
\texttt{BPL3P} & $1000$ & Madau--Fragos & $6.2^{+24}_{-2.4}$ & $0.89^{+5.3}_{-0.33}$ \\
\texttt{symBPL2P} & $300$ & Fitted power law & $6.7^{+12}_{-2.6}$ & $0.89^{+1.2}_{-0.32}$ \\
\texttt{symBPL2P} & $300$ & Madau--Fragos & $6.5^{+12}_{-2.6}$ & $0.88^{+1.2}_{-0.32}$ \\
\texttt{symBPL2P} & $1000$ & Madau--Fragos & $8^{+65}_{-3.9}$ & $1.1^{+16}_{-0.56}$ \\
\texttt{symBPL3P} & $300$ & Fitted power law & $7.4^{+37}_{-3.2}$ & $0.97^{+3.6}_{-0.38}$ \\
\texttt{symBPL3P} & $300$ & Madau--Fragos & $7.2^{+35}_{-3.2}$ & $0.96^{+3.4}_{-0.38}$ \\
\texttt{symBPL3P} & $1000$ & Madau--Fragos & $14^{+3500}_{-9.4}$ & $2.5^{+1500}_{-1.9}$ \\
\texttt{BPL2P} (free slope) & $1000$ & Madau--Fragos & $6.9^{+43}_{-3}$ & $1.1^{+16}_{-0.49}$ \\
\enddata
\tablecomments{The GWTC-3 upper mass cutoff is inferred and differs between
posterior draws. The GWTC-5.0 (LVK) row uses the released posterior;
\texttt{BPL2P} and \texttt{BPL3P} use our primary-mass fits, and \texttt{symBPL2P} and \texttt{symBPL3P} use our symmetric fits to GWTC-5.0.
The free-slope row uses \texttt{BPL2P}
with an independent slope above $300\msun$.
Rows with fitted power-law evolution integrate over $0\leq z\leq1.9$. Madau--Fragos denotes the fixed rate history in Eq.~(\ref{eq:mf17}).
The extrapolations leave the merger rate for every component-mass
pair with $m_1\leq300\msun$ unchanged. For each assumed population model, entries give the median
expected count with errors to its 5th and 95th posterior
percentiles. These errors describe uncertainty in the merger
rate and population parameters; they do not give the range of
integer counts in a realized LISA catalog.
The broad upper interval for \texttt{symBPL3P}
depends on the approximations in Appendix~\ref{app:noise}; its upper endpoint is also approximate because of the finite population sample (Appendix~\ref{app:sampling}).}
\end{deluxetable}

\clearpage
\startlongtable
\begin{deluxetable}{lccccc}
\tablewidth{\textwidth}
\tablecaption{Merger timing of LISA detections with $\rho>8$.
Counts and percentage fractions include mergers during the
four-year mission and within the indicated time after its end.
\label{tab:multiband}}
\tablehead{\colhead{Population} & \colhead{Count, $+30$ yr}
& \colhead{During mission (\%)} & \colhead{$+10$ yr (\%)}
& \colhead{$+30$ yr (\%)} & \colhead{$+100$ yr (\%)}}
\startdata
GWTC-3 & $1.8^{+1.7}_{-0.91}$ & $5.8^{+1.5}_{-1.1}$ & $25^{+4.2}_{-3.4}$ & $42^{+5.3}_{-4.5}$ & $67^{+5}_{-5}$ \\
GWTC-5.0 (LVK) & $3.6^{+5.2}_{-1.6}$ & $14^{+16}_{-6}$ & $39^{+22}_{-11}$ & $55^{+19}_{-9.8}$ & $75^{+11}_{-6.6}$ \\
\texttt{BPL2P} & $3.2^{+9.1}_{-1.5}$ & $15^{+30}_{-8}$ & $39^{+33}_{-12}$ & $54^{+26}_{-11}$ & $74^{+16}_{-7.6}$ \\
\texttt{BPL3P} & $3.4^{+24}_{-1.7}$ & $17^{+46}_{-11}$ & $40^{+45}_{-14}$ & $56^{+35}_{-13}$ & $75^{+20}_{-8.3}$ \\
\texttt{symBPL2P} & $5.2^{+65}_{-3.3}$ & $28^{+44}_{-20}$ & $53^{+40}_{-26}$ & $66^{+30}_{-22}$ & $81^{+17}_{-14}$ \\
\texttt{symBPL3P} & $11^{+3500}_{-8.8}$ & $49^{+38}_{-42}$ & $72^{+28}_{-45}$ & $80^{+20}_{-36}$ & $89^{+11}_{-21}$ \\
\texttt{BPL2P} (free slope) & $4.1^{+43}_{-2.3}$ & $23^{+53}_{-15}$ & $47^{+45}_{-19}$ & $61^{+34}_{-16}$ & $78^{+19}_{-10}$ \\
\enddata
\tablecomments{{Entries show median expected counts or merger fractions, with
lower and upper errors enclosing their central $90\%$ credible intervals.}
The GWTC-3 and GWTC-5.0 rows use
the released posteriors. Our primary-mass fits (\texttt{BPL2P}, \texttt{BPL3P}) and symmetric fits (\texttt{symBPL2P}, \texttt{symBPL3P}) are
extrapolated to $1000\msun$; the free-slope row uses
\texttt{BPL2P} with an independent slope above
$300\msun$. These extrapolations use the Madau--Fragos merger-rate history. No ground-based detection cut is applied.}
\end{deluxetable}

\bibliographystyle{aasjournal}
\bibliography{references}

@ARTICLE{MaganaHernandez:2024Bump,
  author        = {{Maga{\~n}a Hern{\'a}ndez}, Ignacio and {Palmese}, Antonella},
  title         = {{Evidence of a new feature in the binary black hole mass distribution at $70\,M_\odot$ from gravitational-wave observations}},
  journal       = {Phys. Rev. D},
  volume        = {111},
  number        = {8},
  pages         = {083031},
  year          = {2025},
  eprint        = {2407.02460},
  archivePrefix = {arXiv},
  primaryClass  = {astro-ph.HE},
  doi           = {10.1103/PhysRevD.111.083031},
}

@ARTICLE{MaganaHernandez:2025GP,
  author        = {{Maga{\~n}a Hern{\'a}ndez}, Ignacio and {Palmese}, Antonella},
  title         = {{Astrophysics informed Gaussian processes for gravitational-wave populations: Evidence for the onset of the pair-instability supernova mass gap}},
  journal       = {arXiv e-prints},
  eprint        = {2508.19208},
  archivePrefix = {arXiv},
  primaryClass  = {astro-ph.HE},
  year          = {2025},
  month         = aug,
}

@ARTICLE{Abbott:2021GWTC3pop,
  author        = {{Abbott}, R. and {Abbott}, T. D. and {Acernese}, F. and others},
  collaboration = {{LIGO Scientific Collaboration, Virgo Collaboration, and KAGRA Collaboration}},
  title         = {{Population of Merging Compact Binaries Inferred Using Gravitational Waves through GWTC-3}},
  journal       = {Phys. Rev. X},
  volume        = {13},
  pages         = {011048},
  year          = {2023},
  eprint        = {2111.03634},
  archivePrefix = {arXiv},
  primaryClass  = {astro-ph.HE},
  doi           = {10.1103/PhysRevX.13.011048},
}

@article{Ginat:2026awh,
    author = "Ginat, Yonadav Barry and Antonini, Fabio and Flanagan, Elizabeth and Gieles, Mark",
    title = "{Second-Generation Mass Peak in the Gravitational-Wave Population as a Probe of Globular Clusters}",
    eprint = "2604.07456",
    archivePrefix = "arXiv",
    primaryClass = "astro-ph.HE",
    month = "4",
    year = "2026"
}

@ARTICLE{Abbott:2025GWTC4pop,
  author        = {{Abac}, A. G. and {Abouelfettouh}, I. and {Acernese}, F. and others},
  collaboration = {{LIGO Scientific Collaboration, Virgo Collaboration, and KAGRA Collaboration}},
  title         = {{GWTC-4.0: Population Properties of Merging Compact Binaries}},
  journal       = {Astrophys. J. Lett.},
  volume        = {1005},
  number        = {2},
  pages         = {L51},
  year          = {2026},
  eprint        = {2508.18083},
  archivePrefix = {arXiv},
  primaryClass  = {astro-ph.HE},
  doi           = {10.3847/2041-8213/ae771e},
}

@ARTICLE{Abac:2025GW231123,
  author        = {{Abac}, A. G. and {Abouelfettouh}, I. and {Acernese}, F. and others},
  collaboration = {{LIGO Scientific Collaboration, Virgo Collaboration, and KAGRA Collaboration}},
  title         = {{GW231123: A Binary Black Hole Merger with Total Mass 190-265 $M_\odot$}},
  journal       = {Astrophys. J. Lett.},
  volume        = {993},
  pages         = {L25},
  year          = {2025},
  eprint        = {2507.08219},
  archivePrefix = {arXiv},
  primaryClass  = {astro-ph.HE},
  doi           = {10.3847/2041-8213/ae0c9c},
}

@ARTICLE{Abbott:2022IMBHO3,
  author        = {{Abbott}, R. and {Abbott}, T. D. and {Acernese}, F. and others},
  collaboration = {{LIGO Scientific Collaboration, Virgo Collaboration, and KAGRA Collaboration}},
  title         = {{Search for intermediate-mass black hole binaries in the third observing run of Advanced LIGO and Advanced Virgo}},
  journal       = {Astron. Astrophys.},
  volume        = {659},
  pages         = {A84},
  year          = {2022},
  eprint        = {2105.15120},
  archivePrefix = {arXiv},
  primaryClass  = {astro-ph.HE},
  doi           = {10.1051/0004-6361/202141452},
}

@ARTICLE{Sesana:2016,
  author        = {{Sesana}, Alberto},
  title         = {{Prospects for Multiband Gravitational-Wave Astronomy after GW150914}},
  journal       = {Phys. Rev. Lett.},
  volume        = {116},
  number        = {23},
  pages         = {231102},
  year          = {2016},
  eprint        = {1602.06951},
  archivePrefix = {arXiv},
  primaryClass  = {gr-qc},
  doi           = {10.1103/PhysRevLett.116.231102},
}

@ARTICLE{Gerosa:2019Multiband,
  author        = {{Gerosa}, Davide and {Ma}, Sizheng and {Wong}, Kaze W.~K. and {Berti}, Emanuele and {O'Shaughnessy}, Richard and {Chen}, Yanbei and {Belczynski}, Krzysztof},
  title         = {{Multiband gravitational-wave event rates and stellar physics}},
  journal       = {Phys. Rev. D},
  volume        = {99},
  number        = {10},
  pages         = {103004},
  year          = {2019},
  eprint        = {1902.00021},
  archivePrefix = {arXiv},
  primaryClass  = {astro-ph.HE},
  doi           = {10.1103/PhysRevD.99.103004},
}

@ARTICLE{Moore:2019Quiet,
  author        = {{Moore}, Christopher J. and {Gerosa}, Davide and {Klein}, Antoine},
  title         = {{Are stellar-mass black-hole binaries too quiet for LISA?}},
  journal       = {Mon. Not. R. Astron. Soc.},
  volume        = {488},
  number        = {1},
  pages         = {L94--L98},
  year          = {2019},
  eprint        = {1905.11998},
  archivePrefix = {arXiv},
  primaryClass  = {astro-ph.HE},
  doi           = {10.1093/mnrasl/slz104},
}

@ARTICLE{Buscicchio:2025SOBBH,
  author        = {{Buscicchio}, Riccardo and {Torrado}, Jesus and {Caprini}, Chiara and {Nardini}, Germano and {Karnesis}, Nikolaos and {Pieroni}, Mauro and {Sesana}, Alberto},
  title         = {{Stellar-mass black-hole binaries in LISA: characteristics and complementarity with current-generation interferometers}},
  journal       = {J. Cosmology Astropart. Phys.},
  volume        = {2025},
  number        = {01},
  pages         = {084},
  year          = {2025},
  eprint        = {2410.18171},
  archivePrefix = {arXiv},
  primaryClass  = {astro-ph.HE},
  doi           = {10.1088/1475-7516/2025/01/084},
}

@ARTICLE{Babak:2023SGWB,
  author        = {{Babak}, Stanislav and {Caprini}, Chiara and {Figueroa}, Daniel G. and {Karnesis}, Nikolaos and {Marcoccia}, Paolo and {Nardini}, Germano and {Pieroni}, Mauro and {Ricciardone}, Angelo and {Sesana}, Alberto and {Torrado}, Jesus},
  title         = {{Stochastic gravitational wave background from stellar origin binary black holes in LISA}},
  journal       = {J. Cosmology Astropart. Phys.},
  volume        = {2023},
  number        = {08},
  pages         = {034},
  year          = {2023},
  eprint        = {2304.06368},
  archivePrefix = {arXiv},
  primaryClass  = {astro-ph.CO},
  doi           = {10.1088/1475-7516/2023/08/034},
}

@ARTICLE{Farmer:2019MindTheGap,
  author        = {{Farmer}, R. and {Renzo}, M. and {de Mink}, S. E. and {Marchant}, P. and {Justham}, S.},
  title         = {{Mind the Gap: The Location of the Lower Edge of the Pair-instability Supernova Black Hole Mass Gap}},
  journal       = {Astrophys. J.},
  volume        = {887},
  number        = {1},
  pages         = {53},
  year          = {2019},
  eprint        = {1910.12874},
  archivePrefix = {arXiv},
  primaryClass  = {astro-ph.SR},
  doi           = {10.3847/1538-4357/ab518b},
}

@ARTICLE{Gerosa:2021HierMerg,
  author        = {{Gerosa}, Davide and {Fishbach}, Maya},
  title         = {{Hierarchical mergers of stellar-mass black holes and their gravitational-wave signatures}},
  journal       = {Nature Astronomy},
  volume        = {5},
  pages         = {749--760},
  year          = {2021},
  eprint        = {2105.03439},
  archivePrefix = {arXiv},
  primaryClass  = {astro-ph.HE},
  doi           = {10.1038/s41550-021-01398-w},
}

@ARTICLE{Colpi:2024LISARedBook,
  author        = {{Colpi}, Monica and {Danzmann}, Karsten and {Hewitson}, Martin and {Holley-Bockelmann}, Kelly and {Jetzer}, Philippe and {Nelemans}, Gijs and others},
  title         = {{LISA Definition Study Report}},
  journal       = {arXiv e-prints},
  eprint        = {2402.07571},
  archivePrefix = {arXiv},
  primaryClass  = {astro-ph.CO},
  year          = {2024},
  month         = feb,
}

@ARTICLE{Babak:2021SNR,
  author        = {{Babak}, Stanislav and {Hewitson}, Martin and {Petiteau}, Antoine},
  title         = {{LISA Sensitivity and SNR Calculations}},
  journal       = {arXiv e-prints},
  eprint        = {2108.01167},
  archivePrefix = {arXiv},
  primaryClass  = {astro-ph.IM},
  year          = {2021},
  month         = aug,
  note          = {LISA-LCST-SGS-TN-001},
}

@ARTICLE{Wagg:2022LEGWORK,
  author        = {{Wagg}, Tom and {Breivik}, Katelyn and {de Mink}, Selma E.},
  title         = {{LEGWORK: A Python Package for Computing the Evolution and Detectability of Stellar-origin Gravitational-wave Sources with Space-based Detectors}},
  journal       = {Astrophys. J. Suppl. Ser.},
  volume        = {260},
  number        = {2},
  pages         = {52},
  year          = {2022},
  eprint        = {2111.08717},
  archivePrefix = {arXiv},
  primaryClass  = {astro-ph.HE},
  doi           = {10.3847/1538-4365/ac5c52},
}

@ARTICLE{Abbott:2020GW190521,
  author        = {{Abbott}, R. and {Abbott}, T. D. and {Abraham}, S. and others},
  collaboration = {{LIGO Scientific Collaboration and Virgo Collaboration}},
  title         = {{GW190521: A Binary Black Hole Merger with a Total Mass of $150\,M_\odot$}},
  journal       = {Phys. Rev. Lett.},
  volume        = {125},
  number        = {10},
  pages         = {101102},
  year          = {2020},
  eprint        = {2009.01075},
  archivePrefix = {arXiv},
  primaryClass  = {astro-ph.HE},
  doi           = {10.1103/PhysRevLett.125.101102},
}

@ARTICLE{Abbott:2023GWTC3cat,
  author        = {{Abbott}, R. and {Abbott}, T. D. and {Acernese}, F. and others},
  collaboration = {{LIGO Scientific Collaboration, Virgo Collaboration, and KAGRA Collaboration}},
  title         = {{GWTC-3: Compact Binary Coalescences Observed by LIGO and Virgo During the Second Part of the Third Observing Run}},
  journal       = {Phys. Rev. X},
  volume        = {13},
  pages         = {041039},
  year          = {2023},
  eprint        = {2111.03606},
  archivePrefix = {arXiv},
  primaryClass  = {gr-qc},
  doi           = {10.1103/PhysRevX.13.041039},
}

@ARTICLE{Robson:2019LISASens,
  author        = {{Robson}, Travis and {Cornish}, Neil J. and {Liu}, Chang},
  title         = {{The construction and use of LISA sensitivity curves}},
  journal       = {Class. Quantum Grav.},
  volume        = {36},
  number        = {10},
  pages         = {105011},
  year          = {2019},
  eprint        = {1803.01944},
  archivePrefix = {arXiv},
  primaryClass  = {astro-ph.IM},
  doi           = {10.1088/1361-6382/ab1101},
}

@ARTICLE{Cornish:2017GalacticBinary,
  author        = {{Cornish}, Neil J. and {Robson}, Travis},
  title         = {{Galactic binary science with the new LISA design}},
  journal       = {J. Phys. Conf. Ser.},
  volume        = {840},
  number        = {1},
  pages         = {012024},
  year          = {2017},
  eprint        = {1703.09858},
  archivePrefix = {arXiv},
  primaryClass  = {astro-ph.IM},
  doi           = {10.1088/1742-6596/840/1/012024},
}

@MISC{Bradbury:2018JAX,
  author        = {{Bradbury}, James and {Frostig}, Roy and {Hawkins}, Peter and {Johnson}, Matthew James and {Katariya}, Yash and {Leary}, Chris and {Maclaurin}, Dougal and {Necula}, George and {Paszke}, Adam and {Vander{P}las}, Jake and {Wanderman-{M}ilne}, Skye and {Zhang}, Qiao},
  title         = {{JAX: composable transformations of Python+NumPy programs}},
  year          = {2018},
  howpublished  = {\url{https://github.com/jax-ml/jax}},
}

@ARTICLE{Madau:2017SFH,
  author        = {{Madau}, Piero and {Fragos}, Tassos},
  title         = {{Radiation Backgrounds at Cosmic Dawn: X-Rays from Compact Binaries}},
  journal       = {Astrophys. J.},
  volume        = {840},
  number        = {1},
  pages         = {39},
  year          = {2017},
  eprint        = {1606.07887},
  archivePrefix = {arXiv},
  primaryClass  = {astro-ph.HE},
  doi           = {10.3847/1538-4357/aa6af9},
}

@ARTICLE{Abbott:2025GWTC5pop,
  author        = {{Abac}, A. G. and {Abe}, A. and {Abouelfettouh}, I. and others},
  collaboration = {{LIGO Scientific Collaboration, Virgo Collaboration, and KAGRA Collaboration}},
  title         = {{GWTC-5.0: Population Properties of Merging Compact Binaries}},
  journal       = {arXiv e-prints},
  eprint        = {2605.27226},
  archivePrefix = {arXiv},
  primaryClass  = {astro-ph.HE},
  year          = {2026},
  month         = may,
}

@ARTICLE{Reitze:2019CE,
  author        = {{Reitze}, David and others},
  title         = {{Cosmic Explorer: The U.S. Contribution to Gravitational-Wave Astronomy beyond LIGO}},
  journal       = {Bull. Am. Astron. Soc.},
  volume        = {51},
  number        = {7},
  pages         = {035},
  year          = {2019},
  eprint        = {1907.04833},
  archivePrefix = {arXiv},
  primaryClass  = {astro-ph.IM},
}

@ARTICLE{Husa:2016IMRPhenomD,
  author        = {{Husa}, Sascha and {Khan}, Sebastian and {Hannam}, Mark and {P{\"u}rrer}, Michael and {Ohme}, Frank and {Jim{\'e}nez Forteza}, Xisco and {Boh{\'e}}, Alejandro},
  title         = {{Frequency-domain gravitational waves from nonprecessing black-hole binaries. I. New numerical waveforms and anatomy of the signal}},
  journal       = {Phys. Rev. D},
  volume        = {93},
  pages         = {044006},
  year          = {2016},
  eprint        = {1508.07250},
  archivePrefix = {arXiv},
  primaryClass  = {gr-qc},
  doi           = {10.1103/PhysRevD.93.044006},
}

@ARTICLE{Khan:2016IMRPhenomD,
  author        = {{Khan}, Sebastian and {Husa}, Sascha and {Hannam}, Mark and {Ohme}, Frank and {P{\"u}rrer}, Michael and {Jim{\'e}nez Forteza}, Xisco and {Boh{\'e}}, Alejandro},
  title         = {{Frequency-domain gravitational waves from nonprecessing black-hole binaries. II. A phenomenological model for the advanced detector era}},
  journal       = {Phys. Rev. D},
  volume        = {93},
  pages         = {044007},
  year          = {2016},
  eprint        = {1508.07253},
  archivePrefix = {arXiv},
  primaryClass  = {gr-qc},
  doi           = {10.1103/PhysRevD.93.044007},
}

@ARTICLE{CutlerFlanagan:1994Parameters,
  author        = {{Cutler}, Curt and {Flanagan}, {\'E}anna E.},
  title         = {{Gravitational waves from merging compact binaries: How accurately can one extract the binary's parameters from the inspiral waveform?}},
  journal       = {Phys. Rev. D},
  volume        = {49},
  pages         = {2658--2697},
  year          = {1994},
  eprint        = {gr-qc/9402014},
  archivePrefix = {arXiv},
  doi           = {10.1103/PhysRevD.49.2658},
}

@ARTICLE{Peters:1964Radiation,
  author        = {{Peters}, P. C.},
  title         = {{Gravitational Radiation and the Motion of Two Point Masses}},
  journal       = {Phys. Rev.},
  volume        = {136},
  pages         = {B1224--B1232},
  year          = {1964},
  doi           = {10.1103/PhysRev.136.B1224},
}

@ARTICLE{TalbotThrane:2018Mass,
  author        = {{Talbot}, Colm and {Thrane}, Eric},
  title         = {{Measuring the Binary Black Hole Mass Spectrum with an Astrophysically Motivated Parameterization}},
  journal       = {Astrophys. J.},
  volume        = {856},
  pages         = {173},
  year          = {2018},
  eprint        = {1801.02699},
  archivePrefix = {arXiv},
  primaryClass  = {astro-ph.HE},
  doi           = {10.3847/1538-4357/aab34c},
}

@ARTICLE{Fishbach:2018Redshift,
  author        = {{Fishbach}, Maya and {Holz}, Daniel E. and {Farr}, Will M.},
  title         = {{Does the Black Hole Merger Rate Evolve with Redshift?}},
  journal       = {Astrophys. J. Lett.},
  volume        = {863},
  pages         = {L41},
  year          = {2018},
  eprint        = {1805.10270},
  archivePrefix = {arXiv},
  primaryClass  = {astro-ph.HE},
  doi           = {10.3847/2041-8213/aad800},
}

@ARTICLE{Farr:2019Selection,
  author        = {{Farr}, Will M.},
  title         = {{Accuracy Requirements for Empirically Measured Selection Functions}},
  journal       = {Research Notes of the AAS},
  volume        = {3},
  pages         = {66},
  year          = {2019},
  eprint        = {1904.10879},
  archivePrefix = {arXiv},
  primaryClass  = {astro-ph.IM},
  doi           = {10.3847/2515-5172/ab1d5f},
}

@ARTICLE{Mandel:2019Selection,
  author        = {{Mandel}, Ilya and {Farr}, Will M. and {Gair}, Jonathan R.},
  title         = {{Extracting distribution parameters from multiple uncertain observations with selection biases}},
  journal       = {Mon. Not. R. Astron. Soc.},
  volume        = {486},
  pages         = {1086--1093},
  year          = {2019},
  eprint        = {1809.02063},
  archivePrefix = {arXiv},
  primaryClass  = {physics.data-an},
  doi           = {10.1093/mnras/stz896},
}

@ARTICLE{ThraneTalbot:2019Inference,
  author        = {{Thrane}, Eric and {Talbot}, Colm},
  title         = {{An introduction to Bayesian inference in gravitational-wave astronomy: Parameter estimation, model selection, and hierarchical models}},
  journal       = {Publ. Astron. Soc. Australia},
  volume        = {36},
  pages         = {e010},
  year          = {2019},
  eprint        = {1809.02293},
  archivePrefix = {arXiv},
  primaryClass  = {astro-ph.IM},
  doi           = {10.1017/pasa.2019.2},
}

@ARTICLE{Astropy:2022,
  author        = {{Astropy Collaboration} and {Price-Whelan}, Adrian M. and {Lim}, Pey Lian and others},
  title         = {{The Astropy Project: Sustaining and Growing a Community-oriented Open-source Project and the Latest Major Release (v5.0) of the Core Package}},
  journal       = {Astrophys. J.},
  volume        = {935},
  pages         = {167},
  year          = {2022},
  eprint        = {2206.14220},
  archivePrefix = {arXiv},
  primaryClass  = {astro-ph.IM},
  doi           = {10.3847/1538-4357/ac7c74},
}

@ARTICLE{Harris:2020NumPy,
  author        = {{Harris}, Charles R. and {Millman}, K. Jarrod and {van der Walt}, St{\'e}fan J. and others},
  title         = {{Array programming with NumPy}},
  journal       = {Nature},
  volume        = {585},
  pages         = {357--362},
  year          = {2020},
  doi           = {10.1038/s41586-020-2649-2},
}

@ARTICLE{Virtanen:2020SciPy,
  author        = {{Virtanen}, Pauli and {Gommers}, Ralf and {Oliphant}, Travis E. and others},
  title         = {{SciPy 1.0: Fundamental Algorithms for Scientific Computing in Python}},
  journal       = {Nature Methods},
  volume        = {17},
  pages         = {261--272},
  year          = {2020},
  doi           = {10.1038/s41592-019-0686-2},
}

@ARTICLE{Hunter:2007Matplotlib,
  author        = {{Hunter}, John D.},
  title         = {{Matplotlib: A 2D Graphics Environment}},
  journal       = {Computing in Science \& Engineering},
  volume        = {9},
  number        = {3},
  pages         = {90--95},
  year          = {2007},
  doi           = {10.1109/MCSE.2007.55},
}

@ARTICLE{Waskom:2021Seaborn,
  author        = {{Waskom}, Michael L.},
  title         = {{seaborn: statistical data visualization}},
  journal       = {Journal of Open Source Software},
  volume        = {6},
  number        = {60},
  pages         = {3021},
  year          = {2021},
  doi           = {10.21105/joss.03021},
}

@ARTICLE{Abbott:2021GWTC2pop,
  author        = {{Abbott}, R. and others},
  collaboration = {{LIGO Scientific Collaboration and Virgo Collaboration}},
  title         = {{Population Properties of Compact Objects from the Second LIGO--Virgo Gravitational-Wave Transient Catalog}},
  journal       = {Astrophys. J. Lett.},
  volume        = {913},
  number        = {1},
  pages         = {L7},
  year          = {2021},
  eprint        = {2010.14533},
  archivePrefix = {arXiv},
  primaryClass  = {astro-ph.HE},
  doi           = {10.3847/2041-8213/abe949},
}

@ARTICLE{Pierra:2026HeavyBH,
  author        = {{Pierra}, Gr{\'e}goire and {Papadopoulos}, Alexander},
  title         = {{Heavy Black-Holes Also Matter in Standard Siren Cosmology}},
  journal       = {arXiv e-prints},
  year          = {2026},
  eprint        = {2601.03257},
  archivePrefix = {arXiv},
  primaryClass  = {astro-ph.CO},
  doi           = {10.48550/arXiv.2601.03257},
}

@ARTICLE{TiwariFairhurst:2021Structure,
  author        = {{Tiwari}, Vaibhav and {Fairhurst}, Stephen},
  title         = {{The Emergence of Structure in the Binary Black Hole Mass Distribution}},
  journal       = {Astrophys. J. Lett.},
  volume        = {913},
  number        = {2},
  pages         = {L19},
  year          = {2021},
  eprint        = {2011.04502},
  archivePrefix = {arXiv},
  primaryClass  = {gr-qc},
  doi           = {10.3847/2041-8213/abfbe7},
}

@ARTICLE{Antonini:2025ClusterPopulation,
  author        = {{Antonini}, Fabio and {Romero-Shaw}, Isobel and {Callister}, Thomas},
  title         = {{A Star Cluster Population of High Mass Black Hole Mergers in Gravitational Wave Data}},
  journal       = {Phys. Rev. Lett.},
  volume        = {134},
  number        = {1},
  pages         = {011401},
  year          = {2025},
  eprint        = {2406.19044},
  archivePrefix = {arXiv},
  primaryClass  = {astro-ph.HE},
  doi           = {10.1103/PhysRevLett.134.011401},
}

@ARTICLE{FishbachHolz:2020Pairing,
  author = {{Fishbach}, Maya and {Holz}, Daniel E.},
  title = {{Picky Partners: The Pairing of Component Masses in Binary Black Hole Mergers}},
  journal = {\apjl},
  year = {2020},
  volume = {891},
  pages = {L27},
  doi = {10.3847/2041-8213/ab7247},
  eprint = {1905.12669},
  archivePrefix = {arXiv},
  primaryClass = {astro-ph.HE}
}

@ARTICLE{Farah:2024TwoOfAKind,
  author = {{Farah}, Amanda M. and {Fishbach}, Maya and {Holz}, Daniel E.},
  title = {{Two of a Kind: Comparing Big and Small Black Holes in Binaries with Gravitational Waves}},
  journal = {\apj},
  year = {2024},
  volume = {962},
  pages = {69},
  doi = {10.3847/1538-4357/ad0558},
  eprint = {2308.05102},
  archivePrefix = {arXiv},
  primaryClass = {astro-ph.HE}
}

@ARTICLE{Speagle:2020Dynesty,
  author = {{Speagle}, Joshua S.},
  title = {{dynesty: a dynamic nested sampling package for estimating Bayesian posteriors and evidences}},
  journal = {\mnras},
  year = {2020},
  volume = {493},
  pages = {3132--3158},
  doi = {10.1093/mnras/staa278},
  eprint = {1904.02180},
  archivePrefix = {arXiv},
  primaryClass = {astro-ph.IM}
}

@ARTICLE{Phinney:2001Background,
  author = {{Phinney}, E. S.},
  title = {{A Practical Theorem on Gravitational Wave Backgrounds}},
  journal = {arXiv e-prints},
  year = {2001},
  eprint = {astro-ph/0108028},
  archivePrefix = {arXiv},
  doi = {10.48550/arXiv.astro-ph/0108028}
}

@article{Planck:2015Cosmo,
  author  = {{Planck Collaboration} and Ade, P. A. R. and
             Aghanim, N. and others},
  title   = {{Planck 2015 results. XIII. Cosmological parameters}},
  journal = {Astronomy \& Astrophysics},
  year    = {2016},
  volume  = {594},
  pages   = {A13},
  doi     = {10.1051/0004-6361/201525830},
  eprint  = {1502.01589},
  archivePrefix = {arXiv}
}

@ARTICLE{Abac:2026GWTC5Cosmo,
  author = {{Abac}, A. G. and {Abe}, A. and {Abouelfettouh}, I. and others},
  collaboration = {{LIGO Scientific Collaboration, Virgo Collaboration, and KAGRA Collaboration}},
  title = {{GWTC-5.0: Constraints on the Cosmic Expansion Rate and Modified Gravitational-wave Propagation}},
  journal = {arXiv e-prints},
  year = {2026},
  eprint = {2605.27227},
  archivePrefix = {arXiv},
  primaryClass = {astro-ph.CO},
  doi = {10.48550/arXiv.2605.27227}
}

@ARTICLE{Buscicchio:2021LISABBHPE,
  author = {{Buscicchio}, Riccardo and {Klein}, Antoine and {Roebber}, Elinore and
            {Moore}, Christopher J. and {Gerosa}, Davide and {Finch}, Eliot and {Vecchio}, Alberto},
  title = {{Bayesian parameter estimation of stellar-mass black-hole binaries with LISA}},
  journal = {Physical Review D},
  year = {2021},
  volume = {104},
  number = {4},
  pages = {044065},
  doi = {10.1103/PhysRevD.104.044065},
  eprint = {2106.05259},
  archivePrefix = {arXiv},
  primaryClass = {astro-ph.HE}
}

@ARTICLE{Abbott:2023OpenData,
  author        = {{Abbott}, R. and {Abe}, H. and {Acernese}, F. and others},
  collaboration = {{LIGO Scientific Collaboration, Virgo Collaboration, and KAGRA Collaboration}},
  title         = {{Open data from the third observing run of LIGO, Virgo, KAGRA and GEO}},
  journal       = {Astrophys. J. Suppl.},
  volume        = {267},
  pages         = {29},
  year          = {2023},
  doi           = {10.3847/1538-4365/acdc9f},
  eprint        = {2302.03676},
  archivePrefix = {arXiv},
  primaryClass  = {gr-qc},
}

@ARTICLE{TalbotGolomb:2023Likelihood,
  author        = {{Talbot}, Colm and {Golomb}, Jacob},
  title         = {{Growing pains: understanding the impact of likelihood uncertainty on hierarchical Bayesian inference for gravitational-wave astronomy}},
  journal       = {Mon. Not. R. Astron. Soc.},
  volume        = {526},
  number        = {3},
  pages         = {3495--3503},
  year          = {2023},
  eprint        = {2304.06138},
  archivePrefix = {arXiv},
  doi           = {10.1093/mnras/stad2968},
}

@ARTICLE{FishbachKalogera:2021TimeDelays,
  author        = {{Fishbach}, Maya and {Kalogera}, Vicky},
  title         = {{The Time Delay Distribution and Formation Metallicity of LIGO-Virgo's Binary Black Holes}},
  journal       = {Astrophys. J. Lett.},
  volume        = {914},
  pages         = {L30},
  year          = {2021},
  eprint        = {2105.06491},
  archivePrefix = {arXiv},
  primaryClass  = {astro-ph.HE},
  doi           = {10.3847/2041-8213/ac05c4},
}

@ARTICLE{Cutler:2019MultibandScience,
  author        = {{Cutler}, Curt and {Berti}, Emanuele and {Jani}, Karan and {Kovetz}, Ely D. and {Randall}, Lisa and {Vitale}, Salvatore and {Wong}, Kaze W. K. and {Holley-Bockelmann}, Kelly and {Larson}, Shane L. and {Littenberg}, Tyson and {McWilliams}, Sean T. and {Mueller}, Guido and {Schnittman}, Jeremy D. and {Shoemaker}, David H. and {Vallisneri}, Michele},
  title         = {{What we can learn from multi-band observations of black hole binaries}},
  journal       = {arXiv e-prints},
  year          = {2019},
  eprint        = {1903.04069},
  archivePrefix = {arXiv},
  primaryClass  = {astro-ph.HE},
}

@ARTICLE{Datta:2021MultibandGR,
  author        = {{Datta}, Sayantani and {Gupta}, Anuradha and {Kastha}, Shilpa and {Arun}, K. G. and {Sathyaprakash}, B. S.},
  title         = {{Tests of general relativity using multiband observations of intermediate mass binary black hole mergers}},
  journal       = {Phys. Rev. D},
  year          = {2021},
  volume        = {103},
  pages         = {024036},
  doi           = {10.1103/PhysRevD.103.024036},
  eprint        = {2006.12137},
  archivePrefix = {arXiv},
  primaryClass  = {gr-qc},
}

@ARTICLE{Rodriguez:2016ClusterBBH,
  author        = {{Rodriguez}, Carl L. and {Chatterjee}, Sourav and {Rasio}, Frederic A.},
  title         = {{Binary black hole mergers from globular clusters: Masses, merger rates, and the impact of stellar evolution}},
  journal       = {Phys. Rev. D},
  year          = {2016},
  volume        = {93},
  pages         = {084029},
  doi           = {10.1103/PhysRevD.93.084029},
  eprint        = {1602.02444},
  archivePrefix = {arXiv},
  primaryClass  = {astro-ph.HE},
}

@ARTICLE{Antonini:2023ClusterMasses,
  author        = {{Antonini}, Fabio and {Gieles}, Mark and {Dosopoulou}, Fani and {Chattopadhyay}, Debatri},
  title         = {{Coalescing black hole binaries from globular clusters: mass distributions and comparison to gravitational wave data from GWTC-3}},
  journal       = {Mon. Not. R. Astron. Soc.},
  year          = {2023},
  volume        = {522},
  number        = {1},
  pages         = {466--476},
  doi           = {10.1093/mnras/stad972},
}

@ARTICLE{Torniamenti:2024Pairing,
  author        = {{Torniamenti}, Stefano and {Mapelli}, Michela and {P{\'e}rigois}, Carole and {Arca Sedda}, Manuel and {Artale}, M. Celeste and {Dall'Amico}, Marco and {Vaccaro}, M. Paola},
  title         = {{Hierarchical binary black hole mergers in globular clusters: mass function and evolution with redshift}},
  journal       = {Astron. Astrophys.},
  year          = {2024},
  volume        = {688},
  pages         = {A148},
  doi           = {10.1051/0004-6361/202449272},
  eprint        = {2401.14837},
  archivePrefix = {arXiv},
  primaryClass  = {astro-ph.HE},
}

@ARTICLE{Breivik:2016FormationChannels,
  author        = {{Breivik}, Katelyn and {Rodriguez}, Carl L. and {Larson}, Shane L. and {Kalogera}, Vassiliki and {Rasio}, Frederic A.},
  title         = {{Distinguishing Between Formation Channels for Binary Black Holes with LISA}},
  journal       = {Astrophys. J. Lett.},
  year          = {2016},
  volume        = {830},
  pages         = {L18},
  doi           = {10.3847/2041-8205/830/1/L18},
  eprint        = {1606.09558},
  archivePrefix = {arXiv},
  primaryClass  = {astro-ph.GA},
}

@ARTICLE{Nishizawa:2017FormationEccentricity,
  author        = {{Nishizawa}, Atsushi and {Sesana}, Alberto and {Berti}, Emanuele and {Klein}, Antoine},
  title         = {{Constraining stellar binary black hole formation scenarios with eLISA eccentricity measurements}},
  journal       = {Mon. Not. R. Astron. Soc.},
  year          = {2017},
  volume        = {465},
  number        = {4},
  pages         = {4375--4380},
  doi           = {10.1093/mnras/stw2993},
  eprint        = {1606.09295},
  archivePrefix = {arXiv},
  primaryClass  = {astro-ph.HE},
}

@ARTICLE{Vitale:2016Multiband,
  author        = {{Vitale}, Salvatore},
  title         = {{Multiband Gravitational-Wave Astronomy: Parameter Estimation and Tests of General Relativity with Space- and Ground-Based Detectors}},
  journal       = {Phys. Rev. Lett.},
  year          = {2016},
  volume        = {117},
  pages         = {051102},
  doi           = {10.1103/PhysRevLett.117.051102},
  eprint        = {1605.01037},
  archivePrefix = {arXiv},
  primaryClass  = {gr-qc},
}

@ARTICLE{Toubiana:2020LISATestsGR,
  author        = {{Toubiana}, Alexandre and {Marsat}, Sylvain and {Babak}, Stanislav and {Barausse}, Enrico and {Baker}, John},
  title         = {{Tests of general relativity with stellar-mass black hole binaries observed by LISA}},
  journal       = {Phys. Rev. D},
  year          = {2020},
  volume        = {101},
  pages         = {104038},
  doi           = {10.1103/PhysRevD.101.104038},
  eprint        = {2004.03626},
  archivePrefix = {arXiv},
  primaryClass  = {gr-qc},
}

@ARTICLE{GarciaQuiros:2020IMRPhenomXHM,
  author = {{Garc{\'i}a-Quir{\'o}s}, Cecilio and {Colleoni}, Marta
    and {Husa}, Sascha and {Estell{\'e}s}, H{\'e}ctor
    and {Pratten}, Geraint and {Ramos-Buades}, Antoni
    and {Mateu-Lucena}, Maite and {Jaume}, Rafel},
  title = {{IMRPhenomXHM}: A multi-mode frequency-domain model
    for the gravitational wave signal from non-precessing
    black-hole binaries},
  journal = {Phys. Rev. D},
  volume = {102},
  pages = {064002},
  year = {2020},
  doi = {10.1103/PhysRevD.102.064002},
  eprint = {2001.10914},
  archivePrefix = {arXiv},
  primaryClass = {gr-qc}
}

@ARTICLE{Jani:2019IMBHMultiband,
  author        = {{Jani}, Karan and {Shoemaker}, Deirdre and {Cutler}, Curt},
  title         = {{Detectability of intermediate-mass black holes in multiband gravitational wave astronomy}},
  journal       = {Nature Astronomy},
  year          = {2020},
  volume        = {4},
  pages         = {260--265},
  doi           = {10.1038/s41550-019-0932-7},
  eprint        = {1908.04985},
  archivePrefix = {arXiv},
  primaryClass  = {gr-qc}
}

@ARTICLE{FragioneLoeb:2023IMBHHistory,
  author        = {{Fragione}, Giacomo and {Loeb}, Abraham},
  title         = {{Constraining the Cosmic Merger History of Intermediate-mass Black Holes with Gravitational Wave Detectors}},
  journal       = {ApJ},
  year          = {2023},
  volume        = {944},
  pages         = {81},
  doi           = {10.3847/1538-4357/acb34e},
  eprint        = {2212.04056},
  archivePrefix = {arXiv},
  primaryClass  = {astro-ph.HE}
}

@MISC{LVK:2024GWTC3PopulationData,
  author       = {{LIGO Scientific Collaboration} and
                  {Virgo Collaboration} and {KAGRA Collaboration}},
  title        = {{The population of merging compact binaries inferred
                   using gravitational waves through GWTC-3 -- Data release}},
  year         = {2024},
  howpublished = {Zenodo},
  note         = {Version 3},
  doi          = {10.5281/zenodo.11254021},
  url          = {https://zenodo.org/records/11254021}
}

@MISC{LVK:2026GWTC5PopulationData,
  author       = {{LIGO Scientific Collaboration} and
                  {Virgo Collaboration} and {KAGRA Collaboration}},
  title        = {{GWTC-5.0: Population Properties of Merging Compact Binaries}},
  year         = {2026},
  howpublished = {Zenodo},
  note         = {Version 1},
  doi          = {10.5281/zenodo.20292639},
  url          = {https://zenodo.org/records/20292639}
}

@MISC{LVK:2026GWTC5Sensitivity,
  author       = {{LIGO Scientific Collaboration} and
                  {Virgo Collaboration} and {KAGRA Collaboration}},
  title        = {{GWTC-5.0: Cumulative Search Sensitivity Estimates}},
  year         = {2026},
  howpublished = {Zenodo},
  note         = {Version 1},
  doi          = {10.5281/zenodo.19500052},
  url          = {https://zenodo.org/records/19500052}
}

@MISC{LALSuite:2018,
  author       = {{LIGO Scientific Collaboration} and
                  {Virgo Collaboration} and {KAGRA Collaboration}},
  title        = {{LVK Algorithm Library -- LALSuite}},
  howpublished = {Free software (GPL)},
  year         = {2018},
  doi          = {10.7935/GT1W-FZ16}
}

@ARTICLE{Wette:2020SWIGLAL,
  author  = {Wette, Karl},
  title   = {{SWIGLAL}: {Python} and {Octave} interfaces to the
             {LALSuite} gravitational-wave data analysis libraries},
  journal = {SoftwareX},
  volume  = {12},
  pages   = {100634},
  year    = {2020},
  doi     = {10.1016/j.softx.2020.100634}
}

@ARTICLE{Aasi:2015AdvancedLIGO,
  author        = {{Aasi}, J. and others},
  title         = {{Advanced LIGO}},
  year          = {2015},
  journal       = {Class. Quantum Grav.},
  volume        = {32},
  pages         = {074001},
  eprint        = {1411.4547},
  archivePrefix = {arXiv},
  primaryClass  = {gr-qc},
  doi           = {10.1088/0264-9381/32/7/074001},
}

@ARTICLE{Acernese:2015AdvancedVirgo,
  author        = {{Acernese}, F. and others},
  title         = {{Advanced Virgo: a second-generation interferometric gravitational wave detector}},
  year          = {2015},
  journal       = {Class. Quantum Grav.},
  volume        = {32},
  pages         = {024001},
  eprint        = {1408.3978},
  archivePrefix = {arXiv},
  primaryClass  = {gr-qc},
  doi           = {10.1088/0264-9381/32/2/024001},
}

@ARTICLE{Akutsu:2021KAGRA,
  author        = {{Akutsu}, T. and others},
  title         = {{Overview of KAGRA: Detector design and construction history}},
  year          = {2021},
  journal       = {Prog. Theor. Exp. Phys.},
  volume        = {2021},
  pages         = {05A101},
  eprint        = {2005.05574},
  archivePrefix = {arXiv},
  primaryClass  = {physics.ins-det},
  doi           = {10.1093/ptep/ptaa125},
}

@ARTICLE{Toubiana:2022GWTC3LISA,
  author        = {{Toubiana}, Alexandre and {Babak}, Stanislav and {Marsat}, Sylvain and {Ossokine}, Sergei},
  title         = {{Detectability and parameter estimation of GWTC-3 events with LISA}},
  year          = {2022},
  journal       = {Phys. Rev. D},
  volume        = {106},
  pages         = {104034},
  eprint        = {2206.12439},
  archivePrefix = {arXiv},
  primaryClass  = {gr-qc},
  doi           = {10.1103/PhysRevD.106.104034},
}

@ARTICLE{Abbott:2020GW190521Implications,
  author        = {{Abbott}, R. and others},
  title         = {{Properties and Astrophysical Implications of the 150 $M_\odot$ Binary Black Hole Merger GW190521}},
  year          = {2020},
  journal       = {Astrophys. J. Lett.},
  volume        = {900},
  pages         = {L13},
  eprint        = {2009.01190},
  archivePrefix = {arXiv},
  primaryClass  = {astro-ph.HE},
  doi           = {10.3847/2041-8213/aba493},
}

@ARTICLE{Abbott:2019IMBHO1O2,
  author        = {{Abbott}, B. P. and others},
  title         = {{Search for intermediate mass black hole binaries in the first and second observing runs of the Advanced LIGO and Virgo network}},
  year          = {2019},
  journal       = {Phys. Rev. D},
  volume        = {100},
  pages         = {064064},
  eprint        = {1906.08000},
  archivePrefix = {arXiv},
  primaryClass  = {gr-qc},
  doi           = {10.1103/PhysRevD.100.064064},
}

@ARTICLE{LVK:2026GWTC5Intro,
  author        = {{Abac}, A. G. and others},
  title         = {{GWTC-5.0: An Introduction to Version 5.0 of the Gravitational-Wave Transient Catalog}},
  year          = {2026},
  journal       = {arXiv e-prints},
  eprint        = {2605.27223},
  archivePrefix = {arXiv},
  primaryClass  = {gr-qc},
}

@ARTICLE{LVK:2026GWTC5Results,
  author        = {{Abac}, A. G. and others},
  title         = {{GWTC-5.0: Observations from the Second Part of the Fourth LIGO-Virgo-KAGRA Observing Run and Updates to the Gravitational-Wave Transient Catalog}},
  year          = {2026},
  journal       = {arXiv e-prints},
  eprint        = {2605.27225},
  archivePrefix = {arXiv},
  primaryClass  = {gr-qc},
}

@ARTICLE{Essick:2025O4Injections,
  author        = {{Essick}, Reed and {Coughlin}, Michael W. and {Zevin}, Michael and others},
  title         = {{Compact Binary Coalescence Sensitivity Estimates with Injection Campaigns during the LIGO-Virgo-KAGRA Collaborations' Fourth Observing Run}},
  year          = {2025},
  journal       = {Phys. Rev. D},
  volume        = {112},
  pages         = {102001},
  eprint        = {2508.10638},
  archivePrefix = {arXiv},
  primaryClass  = {gr-qc},
  doi           = {10.1103/44x3-hv3y},
}

@ARTICLE{EssickFarr:2022Precision,
  author        = {{Essick}, Reed and {Farr}, Will},
  title         = {{Precision Requirements for Monte Carlo Sums within Hierarchical Bayesian Inference}},
  year          = {2022},
  journal       = {arXiv e-prints},
  eprint        = {2204.00461},
  archivePrefix = {arXiv},
  primaryClass  = {astro-ph.IM},
}

@ARTICLE{HeinzelVitale:2025MonteCarlo,
  author        = {{Heinzel}, Jack and {Vitale}, Salvatore},
  title         = {{When (not) to trust Monte Carlo approximations for hierarchical Bayesian inference}},
  year          = {2025},
  journal       = {arXiv e-prints},
  eprint        = {2509.07221},
  archivePrefix = {arXiv},
  primaryClass  = {astro-ph.IM},
}

@ARTICLE{Evans:2021CEHorizon,
  author        = {{Evans}, Matthew and {Adhikari}, Rana X. and {Afle}, Chaitanya and others},
  title         = {{A Horizon Study for Cosmic Explorer: Science, Observatories, and Community}},
  year          = {2021},
  journal       = {arXiv e-prints},
  eprint        = {2109.09882},
  archivePrefix = {arXiv},
  primaryClass  = {astro-ph.IM},
}

@ARTICLE{Srivastava:2022CETunable,
  author        = {{Srivastava}, Varun and {Davis}, Derek and {Kuns}, Kevin and others},
  title         = {{Science-driven Tunable Design of Cosmic Explorer Detectors}},
  year          = {2022},
  journal       = {Astrophys. J.},
  volume        = {931},
  pages         = {22},
  eprint        = {2201.10668},
  archivePrefix = {arXiv},
  primaryClass  = {astro-ph.IM},
  doi           = {10.3847/1538-4357/ac5f04},
}

@ARTICLE{Klein:2022LastThreeYears,
  author        = {{Klein}, Antoine and {Pratten}, Geraint and {Buscicchio}, Riccardo and others},
  title         = {{The last three years: multiband gravitational-wave observations of stellar-mass binary black holes}},
  year          = {2022},
  journal       = {arXiv e-prints},
  eprint        = {2204.03423},
  archivePrefix = {arXiv},
  primaryClass  = {astro-ph.HE},
}

@ARTICLE{HegerWoosley:2002,
  author        = {{Heger}, Alexander and {Woosley}, S. E.},
  title         = {{The Nucleosynthetic Signature of Population III}},
  year          = {2002},
  journal       = {Astrophys. J.},
  volume        = {567},
  pages         = {532--543},
  eprint        = {astro-ph/0107037},
  archivePrefix = {arXiv},
  primaryClass  = {astro-ph},
  doi           = {10.1086/338487},
}

@ARTICLE{WoosleyHeger:2021Gap,
  author        = {{Woosley}, S. E. and {Heger}, Alexander},
  title         = {{The Pair-Instability Mass Gap for Black Holes}},
  year          = {2021},
  journal       = {Astrophys. J. Lett.},
  volume        = {912},
  pages         = {L31},
  eprint        = {2103.07933},
  archivePrefix = {arXiv},
  primaryClass  = {astro-ph.HE},
  doi           = {10.3847/2041-8213/abf2c4},
}

@ARTICLE{Mangiagli:2019AboveGap,
  author        = {{Mangiagli}, Alberto and {Bonetti}, Matteo and {Sesana}, Alberto and {Colpi}, Monica},
  title         = {{Merger rate of stellar black hole binaries above the pair instability mass gap}},
  year          = {2019},
  journal       = {Astrophys. J. Lett.},
  volume        = {883},
  pages         = {L27},
  eprint        = {1907.12562},
  archivePrefix = {arXiv},
  primaryClass  = {astro-ph.HE},
  doi           = {10.3847/2041-8213/ab3f33},
}

@ARTICLE{GerosaBerti:2019Escape,
  author        = {{Gerosa}, Davide and {Berti}, Emanuele},
  title         = {{Escape speed of stellar clusters from multiple-generation black-hole mergers in the upper mass gap}},
  year          = {2019},
  journal       = {Phys. Rev. D},
  volume        = {100},
  pages         = {041301},
  eprint        = {1906.05295},
  archivePrefix = {arXiv},
  primaryClass  = {astro-ph.HE},
  doi           = {10.1103/PhysRevD.100.041301},
}

@ARTICLE{Racine:2008,
  author        = {{Racine}, {\'E}tienne},
  title         = {{Analysis of spin precession in binary black hole systems including quadrupole-monopole interaction}},
  year          = {2008},
  journal       = {Phys. Rev. D},
  volume        = {78},
  pages         = {044021},
  eprint        = {0803.1820},
  archivePrefix = {arXiv},
  primaryClass  = {gr-qc},
  doi           = {10.1103/PhysRevD.78.044021},
}

@ARTICLE{Ajith:2011,
  author        = {{Ajith}, P. and {Hannam}, M. and {Husa}, S. and others},
  title         = {{Inspiral-merger-ringdown waveforms for black-hole binaries with non-precessing spins}},
  year          = {2011},
  journal       = {Phys. Rev. Lett.},
  volume        = {106},
  pages         = {241101},
  eprint        = {0909.2867},
  archivePrefix = {arXiv},
  primaryClass  = {gr-qc},
  doi           = {10.1103/PhysRevLett.106.241101},
}

@ARTICLE{RouletZaldarriaga:2019,
  author        = {{Roulet}, Javier and {Zaldarriaga}, Matias},
  title         = {{Constraints on Binary Black Hole Populations from LIGO-Virgo Detections}},
  year          = {2019},
  journal       = {Mon. Not. R. Astron. Soc.},
  volume        = {484},
  pages         = {4216--4230},
  eprint        = {1806.10610},
  archivePrefix = {arXiv},
  primaryClass  = {astro-ph.HE},
  doi           = {10.1093/mnras/stz226},
}

@ARTICLE{Miller:2020LowSpin,
  author        = {{Miller}, Simona and {Callister}, Thomas A. and {Farr}, Will M.},
  title         = {{The Low Effective Spin of Binary Black Holes and Implications for Individual Gravitational-Wave Events}},
  year          = {2020},
  journal       = {Astrophys. J.},
  volume        = {895},
  pages         = {128},
  eprint        = {2001.06051},
  archivePrefix = {arXiv},
  primaryClass  = {astro-ph.HE},
  doi           = {10.3847/1538-4357/ab80c0},
}

@ARTICLE{Skilling:2006,
  author        = {{Skilling}, John},
  title         = {{Nested sampling for general Bayesian computation}},
  year          = {2006},
  journal       = {Bayesian Analysis},
  volume        = {1},
  pages         = {833--859},
  doi           = {10.1214/06-BA127},
}

@ARTICLE{Tiwari:2018VT,
  author        = {{Tiwari}, Vaibhav},
  title         = {{Estimation of the sensitive volume for gravitational-wave source populations using weighted Monte Carlo integration}},
  year          = {2018},
  journal       = {Class. Quantum Grav.},
  volume        = {35},
  pages         = {145009},
  eprint        = {1712.00482},
  archivePrefix = {arXiv},
  primaryClass  = {gr-qc},
  doi           = {10.1088/1361-6382/aac89d},
}

@ARTICLE{Hogg:1999,
  author        = {{Hogg}, David W.},
  title         = {{Distance measures in cosmology}},
  year          = {1999},
  journal       = {arXiv e-prints},
  eprint        = {astro-ph/9905116},
  archivePrefix = {arXiv},
  primaryClass  = {astro-ph},
}

@ARTICLE{LVK:2026O4bOpenData,
  author        = {{Abac}, A. G. and others},
  title         = {{Open Data from LIGO, Virgo, and KAGRA through the Second Part of the Fourth Observing Run}},
  year          = {2026},
  journal       = {arXiv e-prints},
  eprint        = {2605.27090},
  archivePrefix = {arXiv},
  primaryClass  = {gr-qc},
}

\end{document}